\documentclass[
  reprint,
  amsmath,
  amssymb,
  nofootinbib,
  prl
]{revtex4-2}
\usepackage{float}
\usepackage{notes2bib}
\usepackage{natbib}[sort&compress]
\usepackage{comment}
\usepackage[english]{babel}
\usepackage[utf8]{inputenc}
\usepackage{graphicx}
\usepackage[colorinlistoftodos, color=green!40, prependcaption]{todonotes}
\usepackage{xr-hyper}
\usepackage[pdftex, pdftitle={Article}, pdfauthor={Author}]{hyperref} 
\usepackage{tabularx}
\usepackage{subcaption}
\usepackage{ragged2e}
\usepackage{diagbox}
\usepackage{caption}
\newcommand{\bra}[1]{\langle #1|}
\newcommand{\ket}[1]{|#1\rangle}
\newcommand{\braket}[1]{\langle #1 \rangle}

\begin{document}
\title{Dissipative Phase Transitions in an Open Quantum Rabi Model with Two-Photon Processes}

\author{Mingjian Zhu}
\email{mz40@rice.edu}
\affiliation{Department of Physics and Astronomy, and Smalley-Curl Institute, Rice University, Houston, TX 77005, USA}
\author{Han Pu}
\email{hpu@rice.edu}
\affiliation{Department of Physics and Astronomy, and Smalley-Curl Institute, Rice University, Houston, TX 77005, USA}
\begin{abstract}
We demonstrate a novel mechanism driving complex critical phenomena in open light-atom interacting systems by investigating a parametrically amplified quantum Rabi model (QRM) subject to both single- and two-photon decay. In the classical oscillator limit, four composite phases emerge, arising from the possible normal or superradiant regimes across the upper and lower spin branches. A mean-field analysis reveals that the two-photon decay activates the intrinsic nonlinearity of the QRM. The synergy of the coherent and dissipative two-photon processes, together with the spin-boson coupling, constitutes an ``inverted" regime where superradiance emerges exclusively at weak coupling. This regime features first- and second-order superradiant DPTs separated by a tricritical point. Utilizing an adiabatic approach and the semi-classical Langevin formalism, we further study the steady-state structure beyond the mean-field level. The universality classes of the DPTs are identified, with the corresponding critical and finite-size scaling exponents derived and a scaling ansatz proposed to describe the critical behavior.
\end{abstract}
\date{\today }
\maketitle
\emph{Introduction} --
The quantum Rabi model (QRM) represents a fundamental paradigm in quantum optics, as it encodes the  light–atom interaction with the minimal set of degrees of freedom \cite{PhysRev.51.652,PhysRevLett.107.100401,Xie_2017}. 
The QRM and its multi-atom extension, the quantum Dicke model (QDM), are well-known for exhibiting the second-order superradiant quantum phase transition \cite{PhysRevA.7.831,PhysRevLett.115.180404}. Significant attention has also been paid to dissipative phase transitions (DPTs), a non-equilibrium generalization of quantum phase transitions, characterized by non-analytical changes in the steady state \cite{PhysRevLett.49.681,PhysRevLett.94.047201,PhysRevA.86.012116,PhysRevA.95.012128,RevModPhys.97.025004,PhysRevA.98.042118}. In addition to their intrinsic physical importance, DPTs can be utilized as robust quantum resources for metrology and error correction applications \cite{PhysRevA.96.013817,PhysRevLett.125.240405,DiCandia2023,PhysRevA.109.022422,sur2025amplifiedresponsecavitycoupledquantumcritical,Petrovnin2025}. Second-order superradiant DPTs have been identified in both open QRM and QDM with single-photon decay \cite{PhysRevA.97.013825,PhysRevA.75.013804}, and have been experimentally observed on various quantum simulation platforms \cite{doi:10.1073/pnas.1417132112,Zhiqiang:17,PhysRevA.97.043858, Chen2021, 10.1063/5.0276914,PhysRevLett.133.173602}. Beyond typical models, theoretical studies of generalized QRMs and QDMs—variants incorporating complex interactions or modified level structures—have revealed a richer landscape of critical phenomena \cite{PhysRevLett.122.193201,PhysRevResearch.6.033075,Zhao_2017,Lu_2024,Wang_2025,azizi2025kerrinducedsuperradianttricriticalpoint,PhysRevA.108.033706,https://doi.org/10.48550/arxiv.2603.13030}. {Specifically, parametric amplification (PA)-a coherent two-photon process commonly used for squeezing generation, photon detection, and quantum gate acceleration \cite{PhysRevA.30.1386,PhysRevLett.57.2520,PRXQuantum.5.020342,10919223,PhysRevLett.122.030501}-is known for driving first-order transitions and tricriticality  \cite{PhysRevLett.124.073602,Yang_2023}.}

Although prior research on these generalized models focuses primarily on coherent interactions, the influence of dissipative mechanisms on DPTs has emerged as a subject of growing interest \cite{PhysRevLett.130.210404,PhysRevLett.133.140403,DiBello2024,nhdh-nvmz,kang2026nongaussianphasetransitioncascade,arxiv.2603.16709}. The state-of-the-art quantum simulation techniques have now enabled the precise engineering of dissipation with high complexity, leading to new regimes of non-equilibrium physics \cite{doi:10.1126/science.1261033,PhysRevLett.118.140403,PhysRevLett.129.233601,So2025,sun2025quantum,doi:10.1126/sciadv.adv7838,so2025experimentalrealizationthermalreservoirs}. In particular, the Markovian two-photon decay, the quantum analog of nonlinear damping in a classical Reyleigh-van der Pol oscillator \cite{thompson2002nonlinear,PhysRevLett.111.234101,PhysRevResearch.3.013130}, exhibits distinctive features of interest \cite{doi:10.1126/science.aaa2085,doi:10.1126/sciadv.ady5649,Beaulieu_2025}.  
Recent studies reveal that it gives rise to interesting physics in light-atom-interacting systems, ranging from parity-dependent non-Hermitian dynamics in the QRM \cite{PhysRevLett.122.043601,PhysRevA.110.023708} to the stabilization of superradiance in the two-photon QDM \cite{mz92-6l9g}. 

 {While the joint effects of coherent and dissipative two-photon processes on DPTs have been extensively studied in purely bosonic systems\cite{PhysRevA.94.033841,PRXQuantum.4.020350,PhysRevResearch.7.013061}, research on light-atom interacting models has typically incorporated only one of these processes. In this paper, we investigate a generalized open QRM that integrates all three components: light-atom interaction, PA, and two-photon decay. We report a novel mechanism driving anomalous superradiance and multicriticality:  the two-photon decay activates the intrinsic nonlinearity of the spin-photon interaction, whose interplay with the two-photon processes generates a rich phase diagram and a new regime of nontrivial physics.} Using mean-field (MF) theory, we show that  the DPTs are extended to both the upper and lower spin branches, yielding four steady-state phases, in contrast to the standard QRM where the transition is restricted to the lower spin branch \cite{PhysRevA.97.013825}. Notably,  we identify an ``inverted" regime in which superradiance emerges when the spin-boson coupling is tuned below a critical value, in stark contrast to the standard QRM/QDM where the superradiant phase is accessed when the coupling is sufficiently large. In the inverted regime, the lower branch supports both first- and second-order DPTs separated by a tricritical point (TCP). We then step beyond the MF theory and obtain an analytic expression of the steady-state Wigner function. Finally, we confirm the tricritical nature of the system by extracting scaling laws for two distinct universality classes and construct a scaling ansatz to describe the finite-size crossover regime.   

\emph{The Model} --
Our Rabi Hamiltonian with PA reads:
\begin{equation}
    H=\frac{\Omega}{2} \sigma_z + \omega_0 a^\dagger a +\frac{\mu}{2}\omega_0(a^2+{a^\dagger}^2)+\lambda\sigma_x(a+a^\dagger),
\end{equation}
where $\sigma_{x,z}$ are spin 1/2 Pauli operators, $a,a^\dagger$ are bosonic ladder operators, and $\mu$ is the normalized strength of PA. The system is subject to one- and two-photon Markovian dissipation, and its dynamics can be described by a Lindblad master equation 
\cite{breuer2002theory}: 
\begin{eqnarray}
        \partial_t\rho=\mathcal{L}\rho&=& -i[H,\rho]
       +2\kappa_1\mathcal{D}_{a}[\rho]
        +2\kappa_2 \mathcal{D}_{a^2}[\rho],
    \label{eq_lind}    
\end{eqnarray}
with $\mathcal{D}_c[\rho]=c\rho c^\dagger+\frac{1}{2}\{c^\dagger c,\rho\}$. Following the notation in Ref.~\cite{PhysRevA.97.013825}, we have denoted the rates of one- and two-photon decay as $2\kappa_1,2\kappa_2$, respectively.

\emph{Mean-field Theory} -- We first solve the steady-state observables using the MF theory, which is precise in the thermodynamic limit (TDL). We renormalize the two-photon decay rate by defining $\kappa_2\equiv\gamma_2\omega_0/\eta$, where $\gamma_2$ is a dimensionless constant and $\eta\equiv\Omega/\omega_0$. Here, the TDL is defined by taking the classical oscillator limit $\eta\rightarrow\infty$ while keeping $\gamma_2$ finite. The steady state then factorizes into the product of a spin state and a coherent bosonic state 
\cite{PhysRevA.97.013825, PhysRevA.94.033841,PhysRevLett.125.240405}. With $\bar{\alpha}=\braket{a}/\sqrt{\eta}=(\bar{x}+i\bar{p})/\sqrt{2}$, after eliminating the spin expectations, the steady-state Heisenberg equations for $\bar{x},\bar{p}$ are given by (see SM, Sec.~\ref*{ap_mf}~\cite{SM_Ref}):
\begin{eqnarray}
    0&=&\left(1-\mu\right)\bar{p}-\gamma_1\bar{x}-\gamma_2\left({\bar{x}}^2+{\bar{p}}^2\right)\bar{x},\nonumber\\
    0&=&-\left(1+\mu\right)\bar{x}\mp f_{\rm nl}(\bar{x})-\gamma_1\bar{p}-\gamma_2\left({\bar{x}}^2+{\bar{p}}^2\right)\bar{p}.
    \label{eq_rabi_mf}
\end{eqnarray}
where $\gamma_1\equiv \kappa_1/\omega_0,\;g\equiv 2\lambda/\sqrt{\omega_0\Omega}$, and we have defined $f_{\rm nl}(x)=g^2x\left(2g^2x^2+1\right)^{-1/2}$ as an effective nonlinear force that encodes the spin-boson coupling. The corresponding steady-state spin polarization takes two different values $s_z=\langle \sigma_z \rangle =\pm (2g^2\bar{x}^2+1)^{-1/2}$, defining two spin branches which we will denote as the upper ($+$) and the lower ($-$) branches. 
The fact that $\gamma_2$ is finite, hence $\kappa_2\propto\eta^{-1}$, ensures that Eq.~\eqref{eq_rabi_mf} remains a valid MF equation independent of $\eta$ \cite{PhysRevLett.133.233604,mz92-6l9g}. In arriving at Eq.~(\ref{eq_rabi_mf}), we have mapped our open QRM to a nonlinear dynamical system.

\begin{figure}[t!]
    \includegraphics[width=1\linewidth]{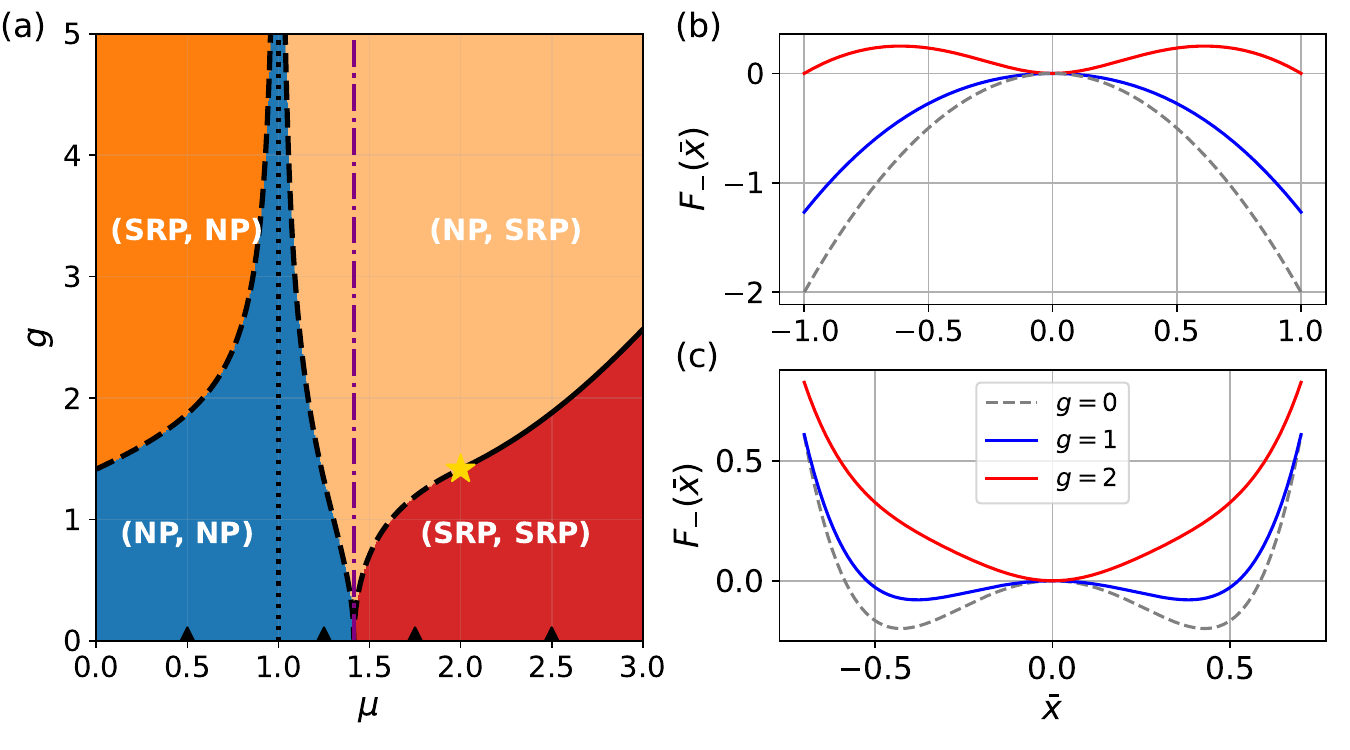}
    \caption{\small \textbf{(a)} Phase diagram in the $(\mu,g)$ parameter space with $\gamma_1=\gamma_2=1$.\footnote{For convenience, we have assumed $\mu\geq0$ and the $\mu<0$ case is discussed in SM, Sec.~\ref*{ap_mf}~\cite{SM_Ref}.} The phases of the $(\pm)$ spin branches are listed in the order (phase($-$), phase($+$)). The dashed (solid) black curves mark the critical condition for DPTs of second- (first-)order. The TCP at $(\mu=2,g=\sqrt{2})$ is highlighted by a yellow star. The purple dot-dashed line at $\mu_c$ marks the boundary of the inverted regime. The dotted line marks the $\mu=1$ case where the two branches stay in NP for all $g$. The four black triangles on the $\mu$ axis mark the selected $\mu$ for plotting Fig.~\ref{fig_mf_n_sz}. In \textbf{(b)} and \textbf{(c)} we plot $F_-(\bar{x})$ without ($\gamma_2=0$) and with ($\gamma_2=1$) nonlinear decay in the inverted regime with $\mu=2,\gamma_1=1$ for different values of $g$.
    }
    \label{fig_phase_diag}
\end{figure}

We introduce the normalized steady-state photon number $\bar{n}_{ss}=\braket{n}(t\to\infty)/\eta$ as the order parameter, with its mean-field value $\bar{n}_{\mathrm{mf}}=|\bar{\alpha}|^2$. $\bar{n}_{\mathrm{mf}} =0$ and $>0$ define normal (NP) and superradiant phase (SRP), respectively. 
To track the emergence of stable non-trivial solutions $(\bar{x}_s,\bar{p}_s)$ of Eq.~\eqref{eq_rabi_mf}, we construct {dissipative functionals $F_{\pm}(x)=\sum_{n=0}^{\infty} c_{2n}^\pm x^{2n}$ \cite{PhysRevA.95.042133}} such that $dF_\pm(\bar{x})/d\bar{x}=0$ yields the equivalent MF equations satisfied by $\bar{x}$ for the two spin branches. 
We shall show later with the Langevin formalism that $F_\pm(x)$ can be mapped to an effective potential. The set of coefficients ${c_{2n}^\pm}$ then allows us to extract critical points of all orders through the Landau theory. {These coefficients also provide direct insight into the mechanism driving the model's exotic critical phenomena.}

\begin{figure}[t!]
\includegraphics[width=1\linewidth]{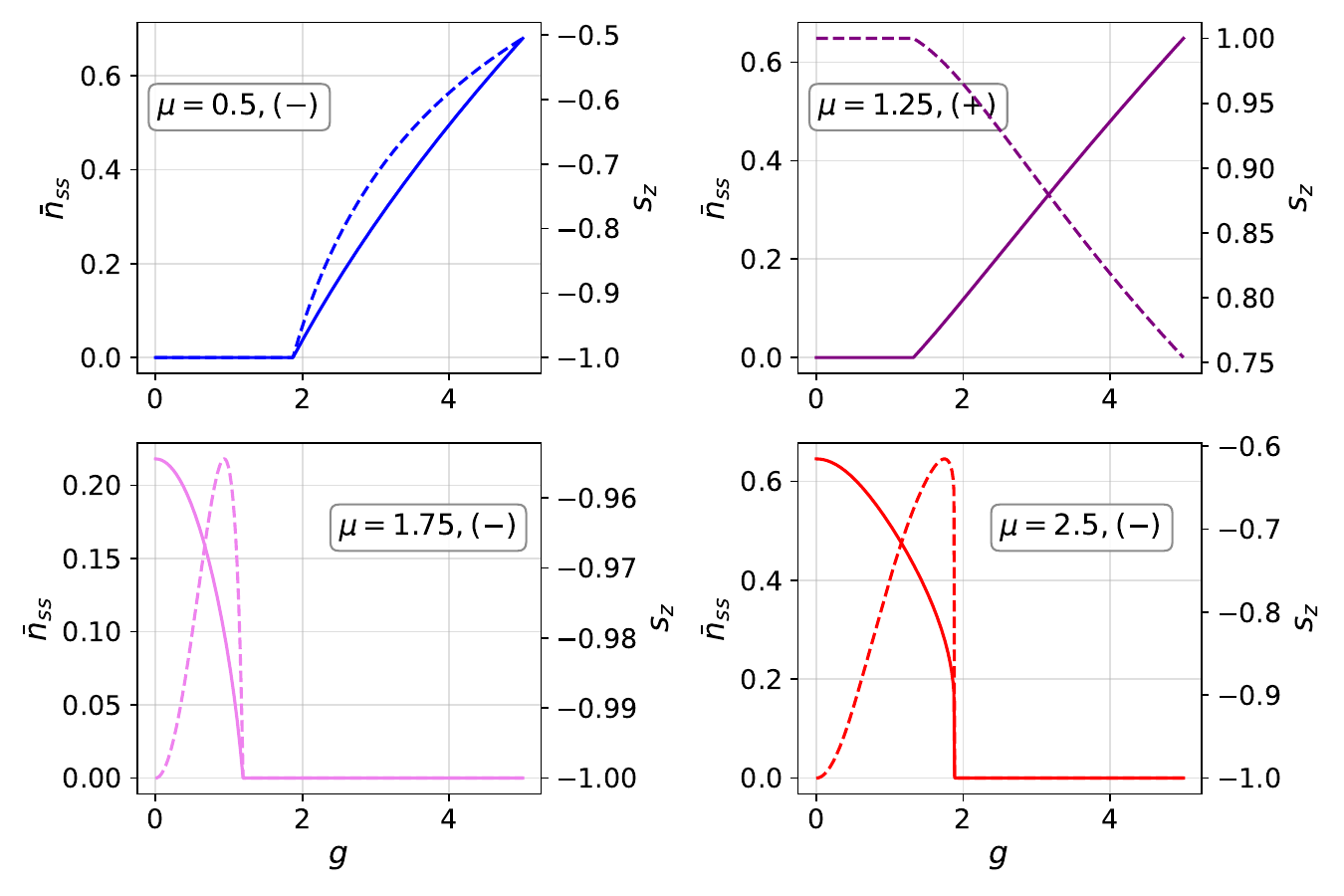}
\caption{\small
Mean-field predictions for the steady-state normalized photon number $\bar{n}_{ss}$  (solid lines, left axis) and spin polarization $s_z$ (dashed lines, right axis) as a function of $g$. The plots illustrate the specific branch undergoing DPT for selected values of $\mu$ (marked by black triangles on the $\mu$ axis of Fig.~\ref{fig_phase_diag}). Other parameters are identical to those in Fig.~\ref{fig_phase_diag}.
}
\label{fig_mf_n_sz}
\end{figure}

The sufficient condition for SRP is obtained by requiring the trivial solution to be unstable, yielding: 
\begin{equation}
  c_2^\pm= (1-\mu)(1+\mu\pm g^2) +\gamma_1^2 < 0.
  \label{eq_phase_cond_1}
\end{equation}
{The presence of PA($\mu$) enables superradiance in the $(+)$ branch, in contrast to the conventional open QRM where the SRP is constrained to the $(-)$ branch.} For the same set of parameters, the two spin branches may exhibit different phases, resulting in a total of four phases, as shown by the example phase diagram in Fig.~\ref{fig_phase_diag}(a).
While the first-order boundary can only be obtained numerically, the second-order boundary is determined analytically by $c_2^\pm=0$, which yields a critical spin-boson coupling:
\begin{equation}
    g_c=\sqrt{|1+\gamma_1^2-\mu^2|/|\mu-1|}.
    \label{eq_gc}
\end{equation}
Some interesting conclusions can be drawn from Eq.\eqref{eq_phase_cond_1},\eqref{eq_gc}: (1) When $\mu=0$, $g_c$ reduces to that of the open QRM with linear decay. As long as $\mu<1$, the $(+)$ branch is always in NP, while the $(-)$ branch is in NP(SRP) when $g<g_c$($g>g_c$), exhibiting a second-order DPT at $g=g_c$. (2) At $\mu=1$ (dotted line in Fig.~\ref{fig_phase_diag}(a)), $g_c\to\infty$ such that both branches remain in the NP for all $g$. The condition $\mu=1$ coincides with the critical point for spectral collapse in a parametrically amplified oscillator \cite{Downing2023}. (3) At $\mu=\mu_c \equiv \sqrt{1+\gamma_1^2}$ (dot-dashed line in Fig.~\ref{fig_phase_diag}(a)), $g_c=0$. It is particularly interesting to note that, when $\mu>\mu_c$, the system enters the ``inverted" regime, where the $(+)$ branch always stays in SRP for all $g\geq0$, and the $(-)$ branch is in SRP(NP) for $g<g_c$($g>g_c$). {This anomalous phenomenon stands in sharp contrast to the conventional QRM, where superradiance emerges only at sufficiently large $g$.} The plots of $F_-(\bar{x})$ in Fig.~\ref{fig_phase_diag}(b) and (c) provide an interpretation of the underlying physics. With $\mu>\mu_c$, the PA induces an anti-confining inverted harmonic potential, whereas the spin-boson coupling acts to restore local stability at $\bar{x}=0$ (see SM, Sec.~\ref*{ap_invert}~\cite{SM_Ref}). In Fig.~\ref{fig_mf_n_sz}, we plot the MF predictions of $\bar{n}_{ss}(g),s_z(g)$ for the branch that undergoes DPT in different regimes throughout the phase diagram.   

{While spin-boson coupling and coherent PA induce anomalous behaviors in the inverted regime, the two-photon decay plays a crucial role in stabilizing the resulting SRPs, as seen by comparing Fig.~\ref{fig_phase_diag}(b) and (c).} Fig.~\ref{fig_phase_diag}(b) shows that $F_-$ does not support stable minimum at $\bar{x} \neq 0$. By contrast, for finite $\gamma_2$(Fig.~\ref{fig_phase_diag}(c)), $F_-$ does possess local minimums at $\bar{x}\neq 0$ when $g<g_c$. A similar stabilization effect has been observed in the two-photon QDM \cite{mz92-6l9g}. {Consequently, the stable realization of the unconventional superradiance is driven by  strict synergy among the three components of the model.}

{The inverted regime is further distinguished by the emergence of tricriticality.} To see this, we shall consider the coefficient $c_4^\pm$ in $F_\pm(x)$: 
\begin{equation}
     c_4^\pm=\pm\left(\mu-1\right)g^4/2+\gamma_2 \gamma_1(2\mu\pm g^2)/(\mu-1).
     \label{eq_c4}
\end{equation}
The sign of $c_4^-$ can be tuned freely when $g=g_c,\mu>\mu_c$, indicating the existence of a TCP in the $(-)$ branch at $c_2^-=c_4^-=0$. With fixed $\mu$ and $\gamma_1$, the tricritical condition can be expressed in terms of a critical nonlinear dissipation rate:
\begin{equation}
\gamma_{2,c}=(\mu-1)^2g_c^4/\left(2\gamma_1(2\mu-g_c^2)\right).
\label{eq_tcp}
\end{equation}
The DPT that occurs in the $(-)$ branch in the inverted regime is of first (second) order for $\gamma_2<\gamma_{2,c}$ ($\gamma_2>\gamma_{2,c}$). {Because only the combined effects of PA ($\mu$), two-photon dissipation ($\gamma_2$), and the spin-boson coupling ($g$) can render $c_2^-,c_4^-$ simultaneously tunable, their confluence constitutes the precise mechanism underlying the tricriticality. A deeper physical interpretation of this mechanism is the explicit activation of the QRM's intrinsic nonlinearity. When $\gamma_2=0$, the MF equation~\eqref{eq_rabi_mf} admits a simple analytical solution exhibiting a standard second-order DPT \cite{PhysRevA.97.013825}. In this regime, only the lowest-order term $c_2^\pm$ in $F_\pm$ is relevant, leaving the higher-order structures in the nonlinear force $f_{\rm nl}$ encoding the spin-boson coupling hidden {and ineffective}. The two-photon decay ($\gamma_2\neq0$), {however}, breaks this analytical simplicity. The additional dissipative nonlinearity makes the higher-order coefficients of $F_{\pm}$ tunable, explicitly promoting the higher-order contributions of $f_{\rm nl}$ into physical relevance. In particular, the TCP emerges precisely because the $o(g^4)$ term of $f_{\rm nl}$ is rendered physically relevant, allowing $c_2^-,c_4^-$ to be tuned to zero simultaneously.}

\emph{Quantum formalism beyond mean field} --
    For realistic experimental and numerical simulations, $\eta$ can only be tuned to a finitely large value. The MF approach then becomes insufficient because it neglects the mixing and non-Gaussian features of the steady state by presuming a simple bosonic coherent state. Here we go beyond the MF by employing the adiabatic approximation: since the energy scale of the spin is much larger than the bosonic frequency, we treat $(a+a^\dagger)=\sqrt 2x$ in the spin-boson interaction as a quasi-static coordinate and diagonalize $H$ in the spin basis through a unitary transformation $U_S = \exp[-i\sigma_y/2\arctan{\left(g(a+a^\dagger )/\sqrt\eta\right)}] $ \cite{PhysRevA.81.042311,Larson_2017}, which decouples the spin and the boson (see SM, Sec.~\ref*{ap_decouple}~\cite{SM_Ref}). After projecting onto the adiabatic spin basis $\ket{\pm}$ (the eigenstates of $U_S^\dag \sigma_z U_S$), we obtain a set of decoupled master equations: 
\begin{eqnarray}
\partial_t\rho_{\pm} &=& -i[H_{\pm},\rho_{\pm}]+2\kappa_1\mathcal{D}_{a}[\rho_{\pm}]+2\kappa_2 \mathcal{D}_{a^2}[\rho_{\pm}],\nonumber\\
H_{\pm}&=&\omega_0 a^\dagger a+ \frac{\mu}{2}\omega_0(a^2+{a^\dagger}^2)\pm V_{\rm nl }(x),
\label{eq_lind_eff}
\end{eqnarray}
where $\rho_{\pm}$ are the bosonic density matrices obtained by projecting $U_S^\dagger\rho U_S$ on $\ket{\pm}$ and we have defined a nonlinear potential: 
\begin{equation}
  V_{\rm nl}(x)=(\omega_0\eta/2)\sqrt{2g^2 x^2/\eta+1},
  \label{eq_V_nl}
\end{equation}
which exactly maps to the nonlinear force in the MF equation as $f_{\rm nl} \propto- dV_n/dx|_{x=\sqrt{\eta}\bar{x}}$, indicating that the adiabatic spin basis $\ket{\pm}$ corresponds to the $(\pm)$ branches defined in the MF theory. {Notably, $V_{\rm nl}$ explicitly manifests the intrinsic nonlinearity of the QRM, as we have demonstrated with the MF approach.}
 
Since there is no coherent coupling between $\ket{+}$ and $\ket{-}$, the steady state in the adiabatic frame can be described by a classical mixture $\rho_{ss}'=p_{+}\rho_{ss,+}\ket{+}\bra{+}+p_{-}\rho_{ss,-}\ket{-}\bra{-}$, where $\rho_{ss,\pm}$ are the bosonic steady states of Eq.~\eqref{eq_lind_eff}. The weights $p_\pm$ are determined by the off-diagonal components of the transformed jump operators in the $\ket{\pm}$ basis \cite{q6xb-8t1j}. In SM, Sec.~\ref*{ap_spin_pop}~\cite{SM_Ref}, which includes Ref. \cite{Zhang_2024}, we derive the expressions of $p_\pm$ perturbatively.  

In the inverted regime, DPTs are confined to the $(-)$ branch. To accurately characterize critical behaviors, it is essential to isolate the $(-)$ branch. Experimentally, this can be achieved by a spin measurement in the adiabatic basis with post selection on $\ket{-}$, which requires an implementation of the unitary transformation $U_S$. In practice, $U_S$ can be approximated by a position-dependent spin rotation $U_{S_1}=\exp(\frac{-ig}{2\sqrt{\eta}}(a+a^\dagger)\sigma_y)$ (see SM, Sec.~\ref*{ap_iso_spin}~\cite{SM_Ref}). Conveniently, $U_{S_1}$ is experimentally accessible via the same techniques employed to generate the spin-boson interaction in the QRM. For the remainder of this work, we focus exclusively on the properties of the $(-)$ branch.  

\begin{figure}[t!]
    \includegraphics[width=0.9\linewidth]{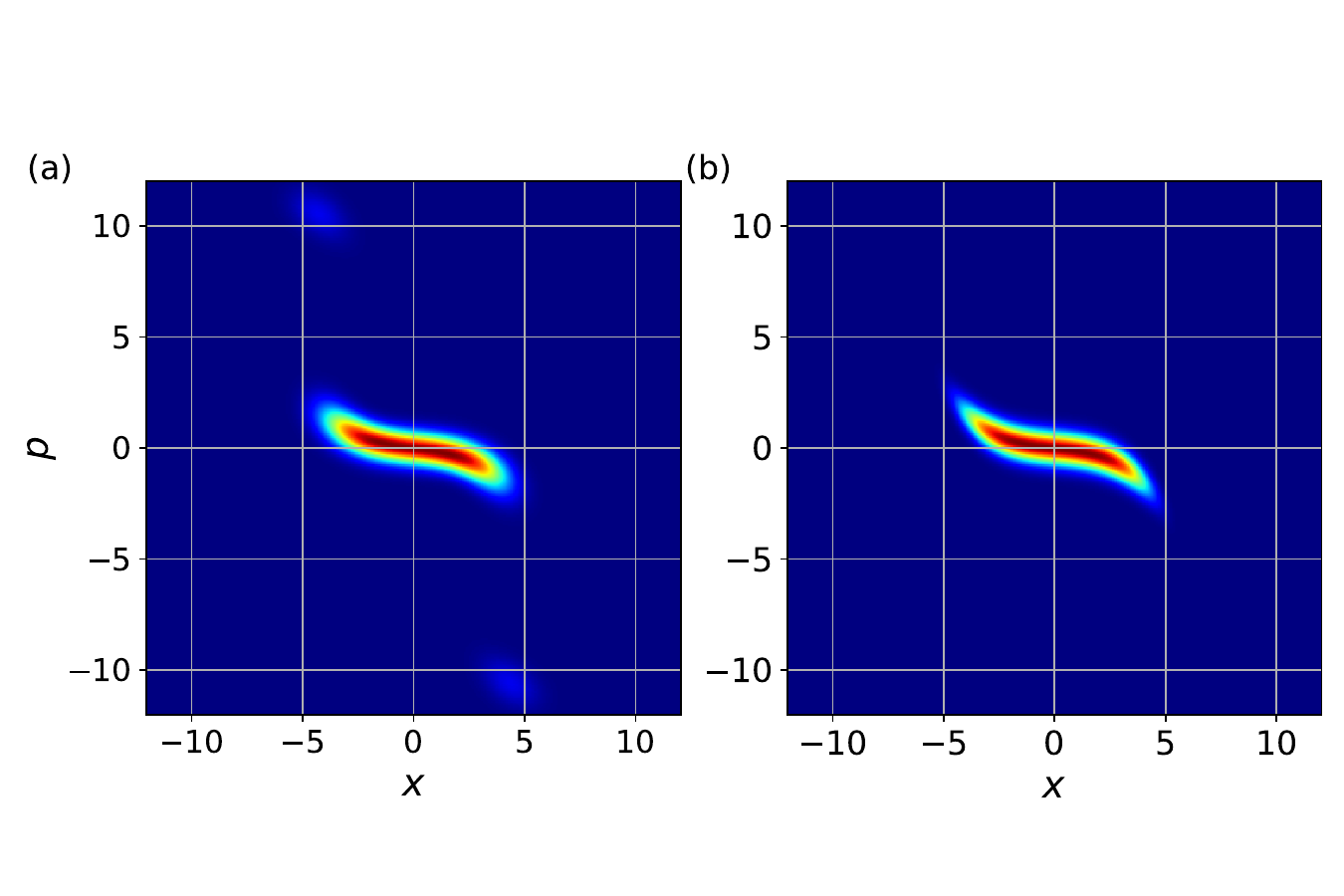}
    \caption{\small \textbf{(a)}. Steady-state Wigner function of the complete model (Eq. \eqref{eq_lind}) calculated numerically. \textbf{(b)}. Steady-state Wigner function of the $(-)$ branch, evaluated using Eq. \eqref{eq_W_ss}. Parameters: $\omega_0=1,\eta=2500,g\sim1.73,$ $\gamma_1=0.1,\gamma_2\sim44.26$. The exact values of $g,\gamma_2$ are determined by the TCP conditions in Eqs. \eqref{eq_gc},\eqref{eq_tcp}. }
    \label{fig_Wigner}
\end{figure} 

\emph{Semi-classical Langevin theory} -- To better understand the properties of $\rho_{ss,-}$ near the TCP, we analyze the system under the semi-classical limit, where $\eta$ is large, but the quantum phase coherence remains significant \cite{Sieberer_2016}. Applying the Keldysh formalism, we map the $(-)$ branch Master equation in \eqref{eq_lind_eff} to a semi-classical Langevin equation by tracking leading-order quantum fluctuations \cite{PhysRevA.87.023831}(See SM, Sec.~\ref*{ap_langevin_derive}, which includes Ref.~\cite{carmichael2013statistical,gardiner2004handbook}) . In the regime of weak linear-dissipation ($\gamma_1 \ll |1-\mu|$), the position variance dominates over the momentum variance ($\Delta x^2\gg\Delta p^2$) and the steady-state Wigner function $W_{ss}^-$ admits an approximate analytical solution which takes a Boltzmann form \cite{PhysRevResearch.7.013061}:
\begin{equation}
    W_{ss}^-=(1/Z_0)\exp{\left[-(mv^2/2+U(x))/T_{\rm eff}\right]},
    \label{eq_W_ss}
\end{equation}
where $Z_0$ is a normalization factor. We define the effective mass as $m=1$ for convenience. With \[v=\partial_t x=-\omega_{0}(\mu-1)p-\kappa_1x-\kappa_2x^3\] and $T_{\rm eff}=\frac{\text{1}}{4}(\mu-1)^{2}$, the effective potential is given by:
\begin{eqnarray}
    U(x)&=&\frac{1}{2}\omega_{0}^{2}[(1-\mu^{2})+\gamma_{1}^{2}]x^{2}+\frac{\omega_{0}^{2}}{2\eta}\gamma_{1}\gamma_{2}x^{4}+\frac{1}{6}\frac{(\omega_{0}\gamma_{2})^{2}}{\eta^{2}}x^{6}\nonumber\\&&+\omega_0(\mu-1)V_{\rm nl}(x)
    \label{eq_eff_variables}
\end{eqnarray} 
where $V_{\rm nl}(x)$ is the nonlinear potential in Eq.~\eqref{eq_V_nl}.

In Fig. \ref{fig_Wigner} we benchmark Eq. \eqref{eq_W_ss} against the steady-state Wigner function of the complete model (Eq.~\eqref{eq_lind}), illustrating that $W_{ss}^-$ successfully captures the structure of the $(-)$ branch (pattern in the center) at the TCP \cite{JOHANSSON20121760}. The faint bimodal structure near the edge of the figure corresponds to the $(+)$ branch in SRP. The relative weights of the two branches can be well explained by the formula of $p_+/p_-$ derived in SM, Sec.~\ref*{ap_spin_pop}~\cite{SM_Ref}. The full Wigner distribution in Fig.~\ref{fig_Wigner}(a) exhibits central symmetry due to the weak $Z_2$ symmetry in Eq.~\eqref{eq_lind}, which enforces $\braket{a}(t\to\infty)=0$ at finite $\eta$. The MF prediction $\bar{\alpha}\neq0$ is only valid in the $\eta\to\infty$ limit with $Z_2$ symmetry breaking\cite{PhysRevA.98.042118,PhysRevLett.125.240405}.

\begin{figure}[t!]
    \includegraphics[width=0.75\linewidth]{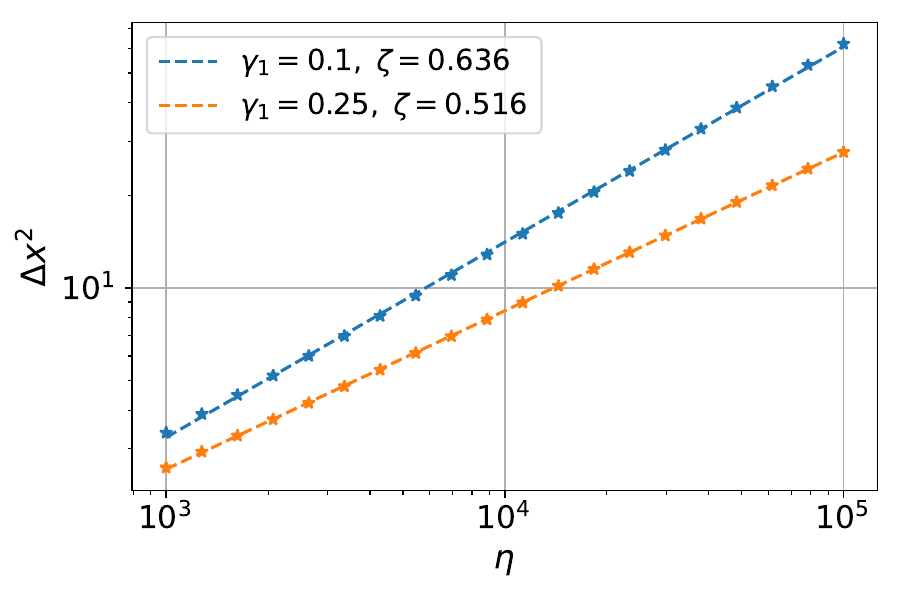}
    \caption{\small Order parameter $\Delta x^2$ as a function of the effective system size $\eta$ at $g=g_c$. The two values of $\gamma_1$ correspond to second-order (orange) and tricritical (blue) DPTs. The dots represent numerical results and the dashed lines are linear fits in log-log space, with extracted slopes ($\zeta_i$) being the estimated finite-frequency scaling exponents. Parameters: $\omega_0=1,\mu=2,$ $\gamma_2\sim44.26$. 
    }
    \label{fig_scaling}
\end{figure}
\emph{Finite-size scaling and universality class} --
A hallmark of tricriticality is the existence of two universality classes, characterized by distinctive scaling laws \cite{nishimori2010elements,goldenfeld2018lectures}. These scaling laws can be correctly predicted by tracking the effective potential $U(x)$ to the order of $x^6$. Expanding $V_{\rm nl}$ to the order of $(\eta^{-2})$ gives:
{\small
\begin{eqnarray}
\label{eq_U_eff_app}
    U(x)&\approx&\sum_{k=1}^3\frac{\omega_0^2}{2k}C_{2k}x^{2k};\;\; C_2=-\left(\mu-1\right)(g_c^2-g^2),\\C_4&=&\frac{1}{\eta}(2\gamma_1\gamma_2- (\mu-1) g^4),\;\; C_6=\frac{1}{\eta^2}(\gamma_2^2+3(\mu-1)g^6/2).\nonumber
\end{eqnarray}
}
Near the critical point, the lower-order coefficients $c_{2n}$ of $F(x)$ defined previously map exactly to $C_{2n}$ in the limit $|1-\mu|\gg\gamma_1$. 

In the regime where Eq. \eqref{eq_W_ss} remains valid, we choose $\Delta x^2$ as the order parameter, as it effectively captures the bosonic fluctuations. The critical exponent $\nu$ can be computed by integrating $W_{ss}^-$ with $g_c\lesssim g$ (see SM, Sec.~\ref*{ap_obs_ss}~\cite{SM_Ref}):
\begin{equation}
\Delta x^2\propto(g-g_c)^{-\nu} = (g-g_c)^{-1},
\label{eq_exp_nu}
\end{equation}

Equation \eqref{eq_exp_nu} is precise when $\eta\to\infty$. The frequency ratio $\eta$ serves as an analog of the effective system size $L$ in a spatially extended system, where the TDL is approached as $L\to \infty$. Consequently, we quantify finite-size effects by analyzing the finite-frequency scaling of the order parameter at the critical coupling ($g=g_c$) \cite{ARDOUREL202399}. With large $\eta$ and $C_4>0$, the $x^4$ term dominates in $U$, while the effect of the $x^6$ term can be observed at the TCP. Integrating $W_{ss}^-$ in the above limits yields: 
\begin{eqnarray} 
\Delta x^2\propto\eta^{\zeta},\;\;\;\zeta = \left\{\begin{array}{ll} 1/2\,,& C_4>0\,\,{\text{(second-order)} } \\ 2/3 \,, & C_4=0 \,\,{\text{(tricritical)} } \end{array} \right.
\label{eq_finite_scale}
\end{eqnarray}  

We show the above scalings in Fig.~\ref{fig_scaling}, where the numerical results are obtained by solving the steady state of the $(-)$ branch decoupled Master equation in \eqref{eq_lind_eff} \cite{note_decouple}. The corresponding coherence length exponents can be obtained as $\xi=2$ for $C_4>0$ and $\xi=3/2$ for $C_4=0$ \cite{PhysRevLett.49.478}. In the Appendix, we summarize these exponents together with those for several different models and propose a scaling ansatz to describe the finite-size crossover behavior near the TCP.

\emph{Conclusion} --
{We have demonstrated a novel mechanism driving complex critical phenomena in the context of an open QRM with coherent and dissipative two-photon processes. With the coherent PA extending superradiance to both spin branches, our MF functional approach reveals that the intrinsic non-linearity of the spin-boson coupling in the QRM is activated by the nonlinear two-photon decay. The synergy of the three components gives rise to an inverted regime where the SRP (NP) emerges for weak (strong) spin-boson coupling, and which also supports a TCP.} An adiabatic treatment followed by the semi-classical Langevin approach further captures the classical mixing and the non-Gaussian structures of the steady state beyond the MF. The approximate analytical solution correctly predicts the two universality classes related to the TCP and the finite-size crossover behaviors. {Our findings demonstrate the crucial roles played by coherent and dissipative two-photon processes in a paradigmatic light-atom-interacting system}. The intrinsic nonlinearity in the QRM revealed by our analysis opens a pathway to higher-order criticality. The infinite-order expansion of $f_{\rm nl}(V_{\rm nl})$ implies that tailored interaction engineering could systematically utilize these higher-order structures to access multicritical points, with the tricriticality we discovered serving as a powerful proof of principle (see SM, Sec.~\ref*{multi_c}~\cite{SM_Ref} for a brief discussion). Exploring these exotic regimes represents a compelling direction for future research.  

\begin{acknowledgments}
The authors acknowledge Yilun Xu for his valuable advice in modeling the system. We acknowledge Hadiseh Alaeian, Diego Fallas Padilla, and Visal So for suggestions and careful reading of the manuscript. We also thank Lin Jiao for insightful discussions.
This work is supported by the NSF (Grant No. PHY-2513089) and the Welch Foundation (Grant No. C-1669).

\emph{Data availability}--Numerical data in this study are available in a public repository\cite{https://doi.org/10.5281/zenodo.19389957}. 
\end{acknowledgments}
\bibliography{refs}{}
\onecolumngrid
\begin{table}[H]
\begin{minipage}{\textwidth}
\centering
\begin{tabular}{|c|c|c|c|c|c|}
\hline
\textbf{Exponent} 
    & \textbf{QRM$^*$ ($C_4>0$)}
    & \textbf{QRM$^*$ (TCP)}
    & \textbf{QRM/QDM(Closed)}
    & \textbf{QRM/QDM $(\kappa_{2}=0)$}
    & \textbf{Kerr/Amplified} \\
\hline
$\nu$   & 1     & 1     & $1/2$ & 1     & 1     \\ \hline
$\zeta$ & $1/2$ & $2/3$ & $1/3$ & $1/2$ & $2/3$ \\ \hline
$\xi$   & 2     & $3/2$ & $3/2$ & 2     & $3/2$ \\ \hline
\end{tabular}
\caption{\small Critical and finite scaling exponents for different models.  In each column we have included the critical exponent for order parameter ($\nu$), the finite scaling exponent ($\zeta$), and the coherence length exponent ($\xi$). From left to right we list the above exponents for the following models: the open QRM studied in this paper ($*$) within the second-order DPT regime ($C_4>0$) and at the TCP ($C_4=0$), the closed QRM/QDM\cite{PhysRevLett.115.180404,https://doi.org/10.1002/qute.201800043}, the open QRM/QDM\cite{PhysRevA.97.013825,PhysRevA.75.013804} with single photon decay, and the resonant Kerr oscillator with single-photon decay/the parametrically amplified oscillator with two-photon decay\cite{PhysRevA.103.033711,PhysRevResearch.7.013061}.}
\label{table_exp}
\end{minipage}
\end{table}
\twocolumngrid
\section*{Appendix}
\label{sec:end_matter}
In Table \ref{table_exp}, we provide a summary of critical, finite-size, and coherence length exponents for the model studied in the main text together with those for several different related models.

\begin{figure}[H]
    \includegraphics[width=0.8\linewidth]{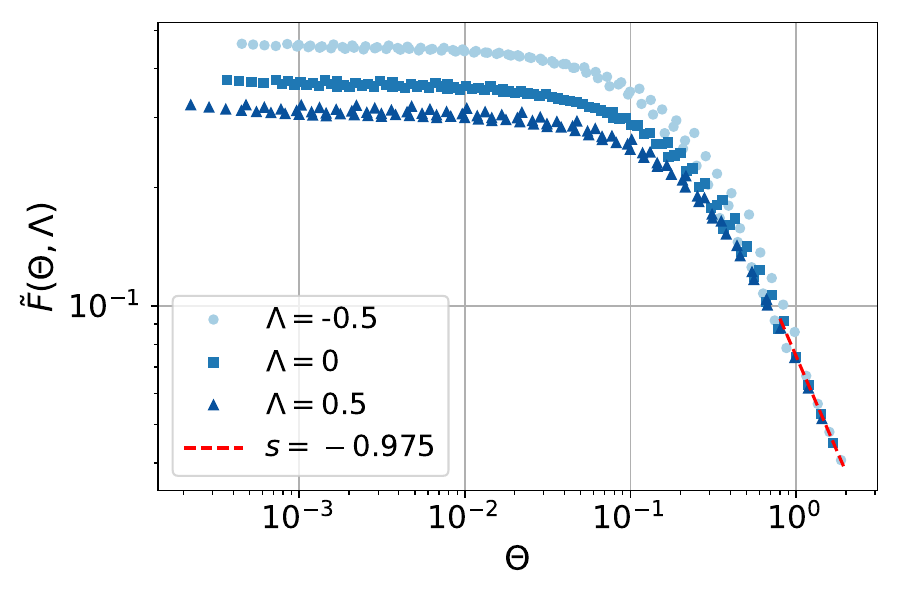}
    \caption{\small Normalized scaling ansatz $\tilde{F}$ as a function of $\Theta$. Data points of different shapes and colors show numerical results calculated for several fixed values of $\Lambda$. The red dashed line represents a linear fit in log-log space performed in the regime $\Theta>0.8$, indicating the corresponding asymptotic power-law behavior described by Eq.~\eqref{eq_exp_nu}. The numerical simulations use fixed parameters  $\omega_0=1,\mu=2,\gamma_1$ and samples from $\eta\in[10^{4},10^{5}]$, $(g-g_c)\in[10^{-5},10^{-2}]$. The value of $\gamma_2$ is determined accordingly such that $\Lambda$ equals to the target fixed value.}
    \label{fig_scaling_F}
\end{figure}

To track the finite-size crossover behaviors near the TCP in the inverted regime, we first set $C_4=0$ in the effective potential (Eq.~\eqref{eq_U_eff_app}) such that the order parameter can be described by the finite-frequency scaling ansatz $\Delta x^2=L^\zeta F_0(\Theta)$ with $L=C_6^{-1/2},\Theta \equiv |g-g_c|L^{\zeta / \nu},\zeta=2/3,\nu=1$ \cite{PhysRevLett.28.1516,binder1992monte}. To account for the effect of $C_4$, we introduce another dimensionless parameter $\Lambda$ by taking $g=g_c$ and unifying the $x^6$ term in $U(x)$ with a change of variable $\tilde{x}=C_6^{1/3}x$. The dependence of $U(x)$ on $C_4,C_6$ is then solely encoded in the factor $\Lambda \equiv C_4L^{4/3}$. With these definitions we can construct a generalized scaling ansatz of the form 
\begin{equation}
    \Delta x^2=L^{2/3}\tilde{F}(\Theta,\Lambda).
    \label{eq_scaling_ansatz}
\end{equation}
{ As an illustrative example, we extract the function $\tilde{F}(\Theta,\Lambda)$ numerically in the inverted regime. We sample a set of $\{\eta,g\}$ near the critical point ($g_c$) for fixed $\gamma_1,\mu,\Lambda$ and solve for  $\gamma_2$ that satisfies the defining equation of $\Lambda$. We numerically evaluated $\Theta$ and the corresponding value of $\Delta x^2 L^{-2/3}$ using the decoupled Master equation \eqref{eq_lind_eff}. 
The results are presented in Fig.~\ref{fig_scaling_F}. For each fixed value of $\Theta$, the collapse of all data points to an approximate single curve validates the ansatz in Eq.\eqref{eq_scaling_ansatz}. $\tilde{F}$ approaches a $\Lambda$–dependent constant as $\Theta\to0$, while for large $\Theta$ it scales as $\tilde{F}\sim\Theta^{-1}$ and becomes independent of $\Lambda$. These limiting behaviors are consistent with asymptotic scalings in Eqs.~\eqref{eq_exp_nu}, \eqref{eq_finite_scale}.}

\end{document}


\title{Supplemantal material to ``Dissipative Phase Transition in a Parametrically Amplified Quantum Rabi Model with Two-photon decay"}

\author{Mingjian Zhu}
\email{mz40@rice.edu}
\affiliation{Department of Physics and Astronomy, and Smalley-Curl Institute, Rice University, Houston, TX 77005, USA}
\author{Han Pu}
\email{hpu@rice.edu}
\affiliation{Department of Physics and Astronomy, and Smalley-Curl Institute, Rice University, Houston, TX 77005, USA}
\maketitle
\section{Determination of critical conditions using the Landau theory}\label{ap_mf}
Following the definitions of $\bar{x},\bar{p}$ in the main text and $s_i=\braket{\sigma_i}$, the corresponding Heisenberg equations in the thermodynamical limit (TDL) are given by:
\begin{eqnarray}
    \partial_t\bar{x}&=&\omega_0[\left(1-\mu\right)\bar{p}-\gamma_1\bar{x}-\gamma_2\left({\bar{x}}^2+{\bar{p}}^2\right)\bar{x}],\nonumber\\
    \partial_t\bar{p}&=&\omega_0[-\left(1+\mu\right)\bar{x}-\frac{g}{\sqrt2}s_x-\gamma_1\bar{p}-\gamma_2\left({\bar{x}}^2+{\bar{p}}^2\right)\bar{p}],\nonumber\\
    \partial_ts_x&=&-\Omega s_y,\;\;\;\partial_ts_z=\Omega\sqrt2{g}\bar{x}s_y,\;\;\;\partial_ts_y=\Omega(s_x-\sqrt2g\bar{x}s_z),
\end{eqnarray}
Setting the LHS to $0$ we obtain the fixed point equations:
\begin{eqnarray}
    0&=&\sqrt2gs_++(1+\mu)\bar{x}+\gamma_1\bar{p}+\gamma_2\bar{p}\left({\bar{x}}^2+{\bar{p}}^2\right),\nonumber\\
    0&=&\left(1-\mu\right)\bar{p}-\gamma_1\bar{x}-\gamma_2\bar{x}\left({\bar{x}}^2+{\bar{p}}^2\right),\nonumber\\
    0&=&-\sqrt2s_++g\bar{x}s_z,\;\;\;
    1=4s_+^2+s_z^2,
    \label{eq_fixed_point_complete}
\end{eqnarray}
where we have introduced $s_+=\braket{\sigma_+}$ and replaced the $s_z$ equation with the spin conservation identity. Solving the last two equations for $s_+$ in terms of $\bar{x}$ and substituting back to the first equation results in the nonlinear effective force $f_{\rm nl}(x)=g^2x\left(2g^2x^2+1\right)^{-1/2}$. The $\bar{x},\bar{p}$ equations can then be combined into a compact form that only involves the variable $u_x=\bar{x}^2$:
\begin{eqnarray}
h\left(u_x\right)&=&u_x(\gamma_1+\gamma_2(u_p+u_x))^2-\left(1-\mu\right)^2u_p=0,\nonumber\\
u_p&=&\frac{1}{1-\mu}u_x[-\left(1+\mu\right)\mp g^2\left(2g^2u_x+1\right)^{-1/2}],
\label{eq_fixed_point_u}
\end{eqnarray}
where we have introduced the variable $u_p=\bar{p}^2$. We note that the $+,-$ signs in $u_p$ arise from the fact that $s_z=\pm (2g^2u_x+1)^{-1/2}$ can take both positive/negative values, which corresponds to the two spin branches $(+,-)$ as mentioned in the main text. 

To obtain the function $F(x)$ in the main text, we expand the nonlinear factor $(2gu_x+1)^{-1/2}$ in $f_{\rm nl}$ into a power series $G(u_x)=\sum_{n=0}^Nb_nu_x^n$, or expanding $f_{\rm nl}(x)$ to the order of $x^{2N+1}$ equivalently. Since we focus on non-trivial solutions, $u_x^1$ can be canceled out from Eq. \eqref{eq_fixed_point_u}, such that the RHS becomes a polynomial of $u_x$ of the order $2N+2$. Integrating  $h(u_x)/u_x$ and replacing $u_x$ by $x^2$ results in  $F_\pm(x)=\sum_{n=0}^{2N+3} c_{2n}^\pm x^{2n}$ in the main text such that $h(u_x)=0$ is equivalent to $dF_\pm(x)/dx=0$. Though $N\to \infty$ is required to precisely describe the solutions of Eq. \eqref{eq_fixed_point_u}, for the purpose of studying tricriticality, it is sufficient to track $c_2^\pm,c_4^\pm$, whose exact forms (Eq. \eqref*{eq_phase_cond_1}, \eqref*{eq_c4} in the main text) can be straightforwardly obtained by expanding $F(x)$ to the order of $N=1$ ($G(u_x)\to1-g^2u_x$).  
The expansion remains a fairly good approximation when $g^2u_x\ll1$, which is valid near the critical point for a second-order dissipative phase transition (DPT). For the first-order DPT in the system, this condition is usually satisfied when $1\lesssim\gamma_2$ suggested by numerical results. 
\begin{figure}[t!]
    \includegraphics[width=0.6\linewidth]{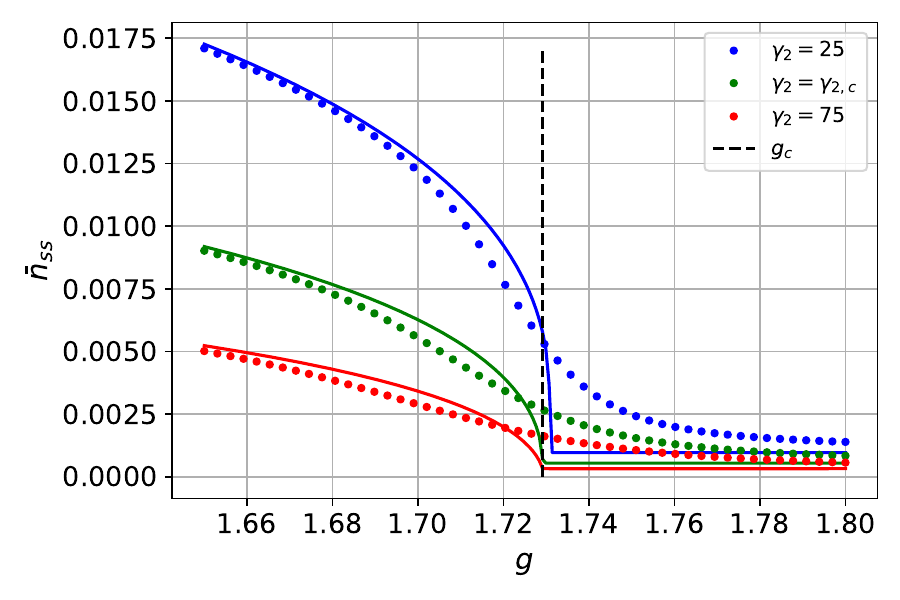}
    \caption{\small Normalized order parameter $\bar{n}_{ss}$ as a function of $g$ for different $\gamma_2$. Dots represent the numerical solutions of the complete model (steady state of Eq. \eqref*{eq_lind} of the main text) while dashed lines show mean-field predictions incorporating the classical mixture ansatz (see Appendix \ref{ap_spin_pop}). Colors denote the nature of the DPT: first-order (blue), tricritical (green), and second-order (red). The black dashed line marks the critical coupling for the second-order DPT. The deviation from MF predictions near the critical point demonstrates finite-size effects\cite{note_finite_eta}. The parameters used for simulations are: $\omega_0=1,\eta=2500,\mu=2,$ $\gamma_1=0.1.$ and the corresponding $\gamma_{2,c}\sim44.26$.
    }
    \label{fig_dpt_sig}
\end{figure}

To obtain the conditions for the emergence of non-trivial solutions, we  consider the exact equation $\tilde{f}(u_x)=H(u_x)/u_x=0$. Since $\tilde{f}(u_x)\to \infty$ as $u_x\to \infty$ (assuming $\gamma_2>0$), if $\tilde{f}(0)=c_2^\pm<0$, there is at least one solution $u_x>0$ with $\tilde{f}'(u_x)>0$ (stable non-trivial solution). Hence, $c_2^\pm<0$ is a sufficient condition for the emergence of superradiant phase(SRP). However, $c_2^\pm=0$ cannot be used as the exact critical condition for DPT, since a first-order DPT can happen when a non-trivial $x$ emerges as a global minimum of $F(x)$, although $F(0)$ remains as a stable local minimum simultaneously, which is a common case in a Landau $\phi^6$ theory. In our model, due to the nonlinearity, the exact critical condition can only be obtained numerically. 

In Fig. \ref{fig_dpt_sig}, we show the transition from first- to second-order DPT across the tricritical point (TCP) by plotting the order parameter $\bar{n}_{ss}$ as a function of $g$ for different nonlinear dissipation strength $\gamma_2$, demonstrating a good consistency between mean field(MF) and numerical results at large $\eta$.  

\begin{figure}[t!]
    \includegraphics[width=0.5\linewidth]{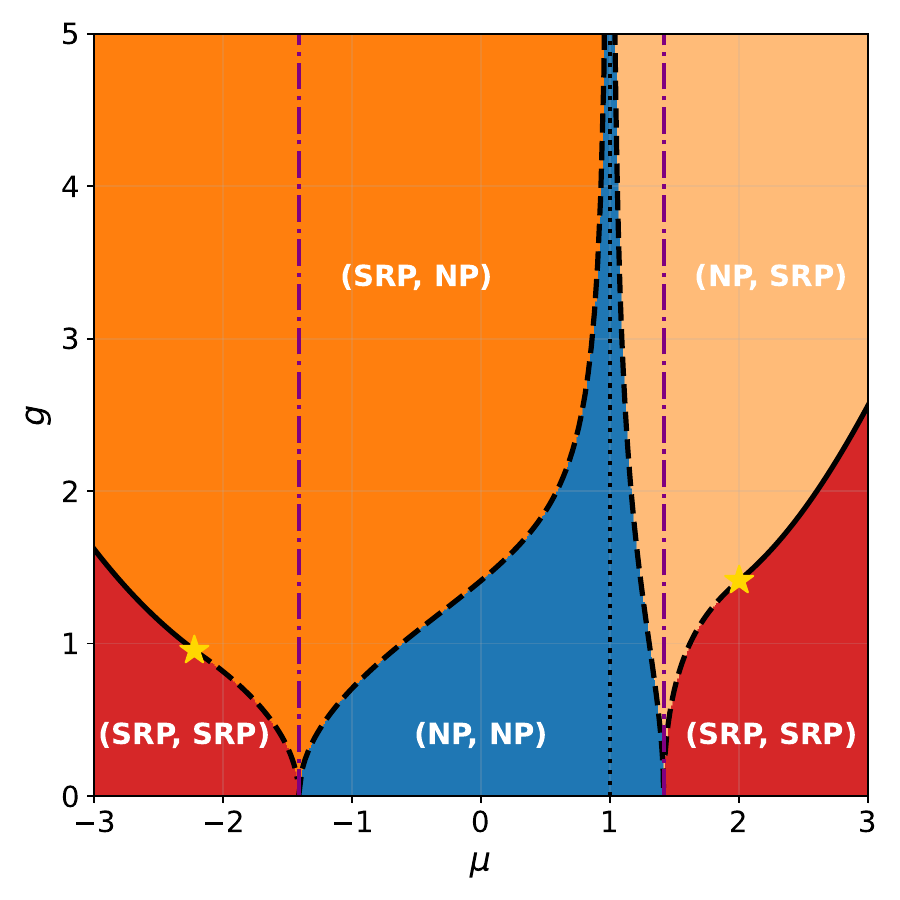}
    \caption{\small \textbf{(a)} Phase diagram in the $(\mu,g)$ parameter space with $\gamma_1=\gamma_2=1$. In each regime the phases of the $(\pm)$ spin branches are listed in the order \{phase ($-$), phase (+)\}. The dashed (solid) black curves marks the critical condition for DPTs of second- (first-)order. The two TCPs are highlighted by yellow stars. The purple dot-dashed lines at $\mu=\pm\mu_c$ mark the boundaries of the two inverted regimes. The dotted line marks the $\mu=1$ case where the two branches stay in NP for all $g$. 
    }
    \label{fig_phase_diag_full}
\end{figure}

{The phase diagram in the main text can be extended to the regime $\mu<0$, where the parametric amplification(PA) term carries an additional phase of $\pi$. The following results can be obtained by analyzing Eq. \eqref*{eq_phase_cond_1},\eqref*{eq_gc} in the main text. (1) When $0>\mu>-\mu_c$, the $(+)$ branch stays in normal phase(NP) while the $(-)$ branch transits to SRP when $g>g_c$. This regime is an extension of the $0<\mu<1$ normal regime. (2) When $\mu<-\mu_c$, the system enters another inverted regime. In contrast to the inverted regime at $\mu>\mu_c$, here the $(-)$ branch always exhibits superradiance due to the PA, even for the case $g=0$. Tricriticality now emerges in the $(+)$ branch, which transits into SRP at sufficiently small $g$. The exact position of the TCP can be determined by requiring $c_2^+=c_4^+=0$. We summarize the above features by presenting the complete phase diagram in Fig. \ref{fig_phase_diag_full}.  }

\section{Stabilization effect of two-photon dissipation}
\label{ap_mf_0}
The nonlinear dissipation guarantees the existence of a stable fixed point by enforcing $\lim_{ u_x\to\infty}\tilde{f}(u_x)\to\infty$. This stabilization is specifically important for  the inverted regime ($\mu>\mu_c=\sqrt{1+\gamma_1^2}$) as $\lim_{ u_x\to\infty}\tilde{f}(u_x)\to1+\gamma_1^2-\mu^2<0$ for $\gamma_2=0$ and it can be further demonstrated that no stable non-trivial fixed point exists without the two-photon decay. 

When $\gamma_2=0$ the  steady-state equation \eqref{eq_fixed_point_complete} yields only an unstable trivial solution $\bar{x}=\bar{p}=0$ for the $(+)$ branch. For the $(-)$ branch, when $g<g_c$ non-trivial solution does not exist, which is very different from the $\gamma_2>0$ case where superradiance emerges for $g<g_c$. A closed form non-trivial solution exists for $g>g_c$ :
\begin{eqnarray}
    s_z=\frac{1+\gamma_1^2-\mu^2}{g^2(\mu-1)}=-\frac{g_c^2}{g^2},\;\;\bar{x}=-\frac{1}{\sqrt2g}\sqrt{1/s_z^2-1},\;\;s_+&=&\frac{1}{\sqrt2}g\bar{x}s_z,\;\; \bar{p}=\frac{\gamma_1}{1-\mu}\bar{x}
    .
    \label{eq_sol_linear}
\end{eqnarray}

However, this solution is not stable, which can be verified by a linear stability analysis. Starting from Eq.\eqref*{eq_rabi_mf} in the main text, we substitute $s_z=-\sqrt{1-s_x^2-s_y^2}$ such that the remaining equations become independent. We then compute the determinant of the corresponding Jacobi matrix at the fixed point given by Eq. \eqref{eq_sol_linear}:
\begin{equation}
    \det{[J]}=(\gamma_1^2+1-\mu^2 )  \frac{1}{s_z}  (s_z+\frac{1}{s_z}).
\end{equation}
Since $s_z<0, \gamma_1^2+1-\mu^2 <0$, $\det [J]<0$, this solution is always unstable. For the trivial solution, following a similar analysis one can show that the related stability condition is given by $c_2^->0$, where $c_2^-$ is defined in Eq. \eqref*{eq_phase_cond_1} in the main text. In the inverted regime this simplifies to $g>g_c$, which means the NP is always stable even without the two-photon decay.

\section{Derivation of the decoupled bosonic master equations}\label{ap_decouple}
We first derive the transformed Rabi Hamiltonian under the adiabatic approximation. The spin-dependent part of the Rabi Hamiltonian can be considered as a typical two-level system with an effective detuning $\Omega$ and Rabi frequency $\lambda(a+a^\dagger)$, which can be diagonalized with a position-dependent rotation 
$U_S = \exp(-i\sigma_y/2\arctan{\left(g(a+a^\dagger\right)/\sqrt\eta)})$. In the adiabatic basis the transformed Hamiltonian takes the form: 
\begin{eqnarray}
    H_{\rm d}=U_S^\dagger HU_S=\sigma_zV_{\rm nl}(x)+\omega_0 a^\dagger a+\frac{\mu}{2}\omega_0(a^2+{a^\dagger}^2),\;\;V_{\rm nl}(x)=\frac{\omega_0\eta}{2}\sqrt{2g^2x^2/\eta+1}.
\end{eqnarray}
where $x=(a+a^\dagger)/\sqrt{2}$. We have neglected the effects of $U_s$ on the purely bosonic part $H_B=\omega_0[a^\dagger a+\mu(a^2+{a^\dagger}^2)]$. The off-diagonal terms arising from $U_S^\dagger H_BU_S$ are at most of the order $o(\eta^0)$, which has a negligible effect compared to the dominant interaction term $V_{\rm nl} \propto \eta^1$, since $\eta$ is large. A series expansion of $V_{\rm nl}$ will reproduce the Schrieffer-Wolff transformation derived in the SM of \cite{PhysRevLett.115.180404} up to arbitrary orders.   

We then apply the transformation on both sides of the complete master equation \eqref*{eq_lind} in the main text and evaluate $U^\dagger_S\mathcal{L}[\rho]U_S$. Projecting $H_{\rm d}$ on the adiabatic spin basis $\ket{\pm}$ leads to the coherent parts of Eq. \eqref*{eq_lind_eff} in the main text. We now argue that the jump operators stay effectively invariant in the  decoupled equations. Defining $\theta(x)=\arctan{(gx/\sqrt{\eta/2})}$ such that $ U_S=\exp(-i\sigma_y\theta(x)/2)$ we have 
\begin{eqnarray}
    c_1=U_S^
    \dagger a U_S=
    a+\frac{i}{\sqrt{2}}(U_S^
    \dagger[p,U_S])=a+\frac{1}{\sqrt{2}}(U_S^\dagger
    \partial_x U_S)=a-i\epsilon(x)\sigma_y, 
    \label{eq_c1_comp}
\end{eqnarray}
where $\epsilon(x)=\frac{1}{2\sqrt{2}}\partial_x\theta(x)$, and 
\begin{equation}
    c_2=U_S^
    \dagger a^2 U_S=a^2-i(\epsilon(x)a+\epsilon(x)a)\sigma_y-\epsilon(x)^2. 
    \label{eq_c2_comp}
\end{equation}
Consider the transformed Lindbladian dissipator $\mathcal{D}_{c_i}[\rho]$. For $c_1$, the correction $\epsilon(x)\sigma_y$ only generates off-diagonal effects through the quantum jump term $c_i\rho c_i^\dagger$ after projecting on $\ket{\pm}$. The only diagonal contribution is induced by the anti-commutator $\{c_i^\dagger c_i,\rho\}$ and is of the order $o(\eta)^{-1}$ since $\epsilon(x)$ is of the order $o(\eta^{-1/2})$ in both NP and SRP. It can be safely neglected when $\eta$ is large. Therefore, to the leading order $c_1\sim a$ in the decoupled equations. A similar argument can be made in terms of projection and scaling to demonstrate that $c_2\sim a^2$.

\section{Perturbative calculation of the spin population}\label{ap_spin_pop}
We aim to derive a perturbative solution for the spin weights $p_{\pm}$ in the classical mixture ansatz $\rho_{ss}'=p_{+}\rho_{ss,+}\ket{+}\bra{+}+p_{-}\rho_{ss,-}\ket{-}\bra{-}$. In the NP, the bosonic excitation is small and does not grow with $\eta$. When $\eta$ is large but finite, the steady state is a unique mixed state due to the effect of the off-diagonal terms in $c_i$. A perturbative treatment performed to the order of $(x^2/\eta)^1$ in \cite{q6xb-8t1j} shows that $p_+=p_-=0.5$ in this case. In the strict limit $\eta\to\infty$, the off-diagonal terms can be completely neglected and the dynamics of the system can be tracked by the decoupled bosonic Master equation \eqref*{eq_lind_eff} in the main text, demonstrating the existence of a  strong spin symmetry in the $\ket{\pm}$ basis as $H,c_i$ all commutes with $\sigma_z$. The steady state is not unique and the weights $p_+,p_-$ are determined by the initial condition\cite{q6xb-8t1j,Zhang_2024}.
\begin{figure}[t!]
    \includegraphics[width=0.75\linewidth]{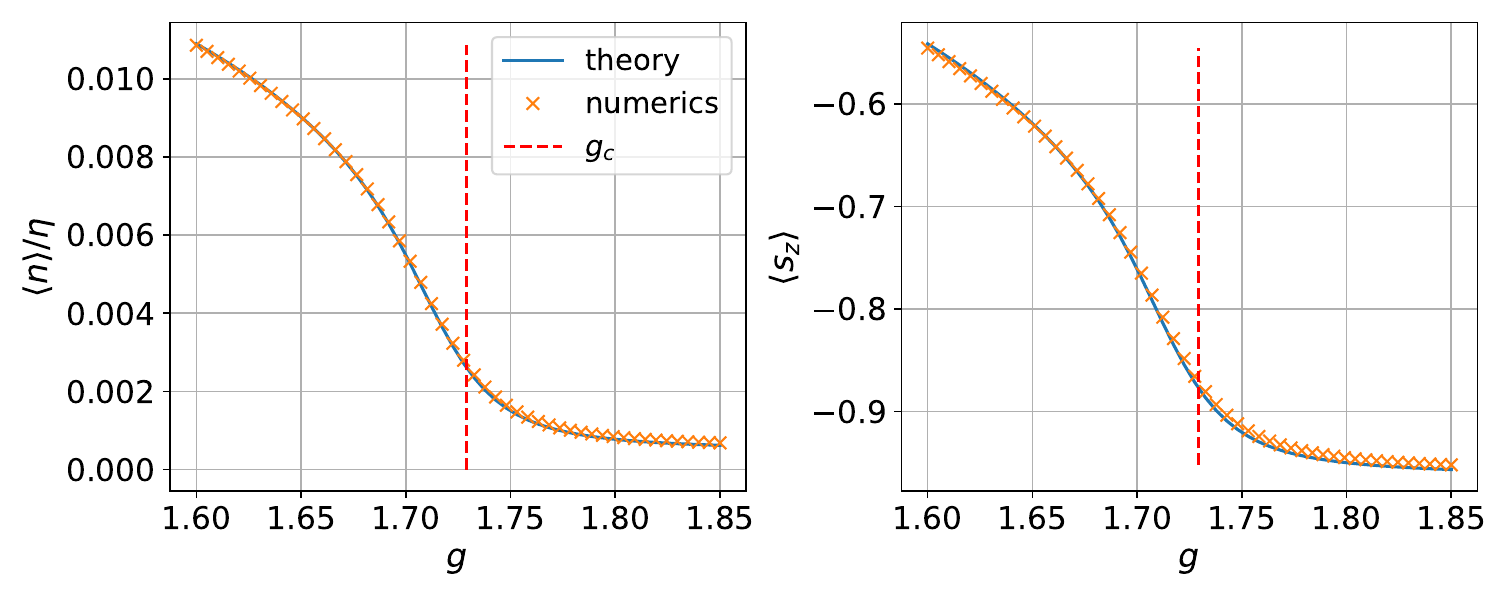}
    \caption{\small Verification of the approximate branch weight relation in Eq. \eqref{eq_spin_pop}. The dots show the steady-state expectation values of $n,s_z$ obtained by numerically solving the complete model. The solid line plots the prediction of the perturbative formalism, which is obtained by first solving the decoupled master equations corresponding to Eq. \eqref{eq_decoupled_6} numerically and then evaluating mixed state expectation values using Eq. \eqref{eq_spin_pop}. The parameters used for simulation are $\eta=2500,\gamma_1=0.1,\omega_0=1,\mu=2$. $\gamma_2$ is set to the TCP point value $\sim44.25$.  The Fock space cutoff $n_c$ is chosen for each $g$ such that $\max({\rm tr}(\rho_{ss,n}\ket{n_c-1}\bra{n_c-1}))\sim 10^{-10}$. 
    }
    \label{fig_spin_pop}
\end{figure} 
\begin{figure}[h!]
    \includegraphics[width=0.75\linewidth]{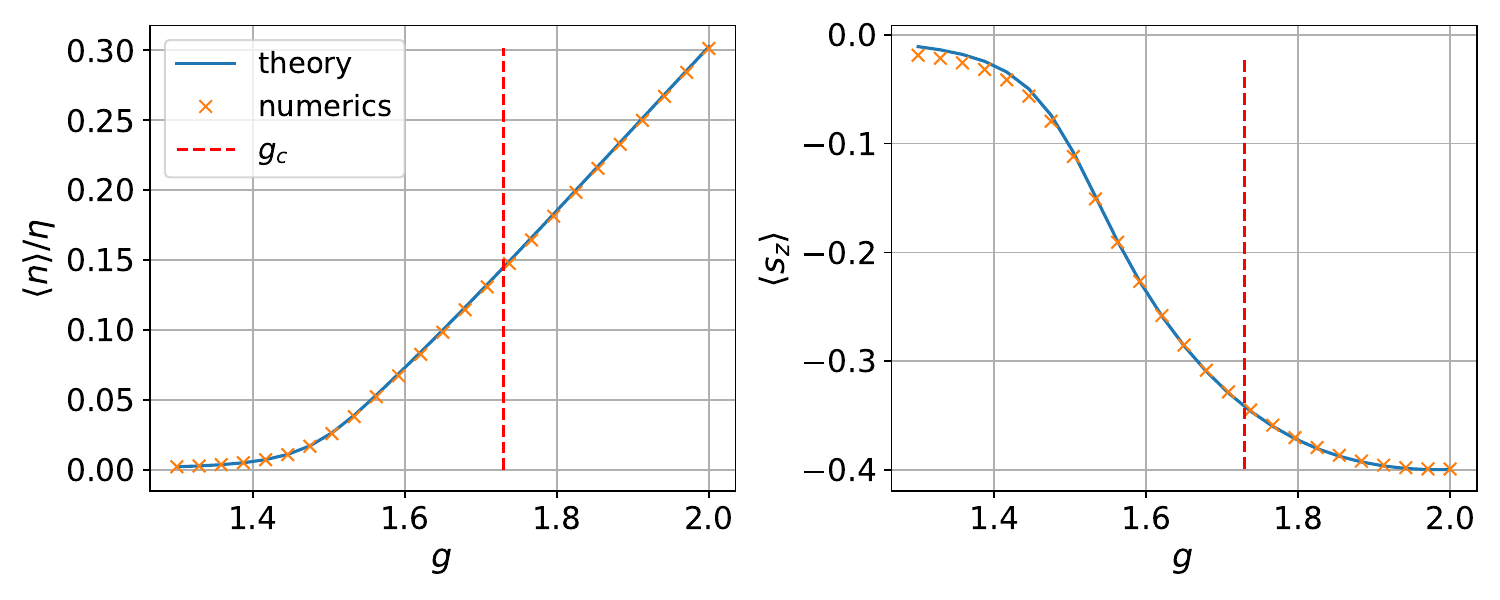}
    \caption{\small Verification of the complete branch weight relation in Eq. \eqref{eq_spin_pop_comp}. The dots show the steady-state expectation values of $n,s_z$ obtained by numerically solving the complete model. The solid line plots the prediction of the perturbative formalism, which is obtained by first solving the decoupled master equations corresponding to Eq. \eqref*{eq_lind_eff} in the main text numerically and then evaluating mixed state expectation values using Eq. \eqref{eq_spin_pop_comp}. The parameters used for simulation are $\eta=250,\omega_0=1,\mu=0,\gamma_1=1,\gamma_2=0.1$. 
    }
    \label{fig_spin_pop_comp}
\end{figure}

In the SRP, the bosonic mode gains excitations proportional to $\eta$ such that the off-diagonal terms in $c_i$ generates non-trivial incoherent effects which can no longer be neglected even in the limit $\eta\to\infty$. Therefore, the steady state remains a unique mixed state for all values of $\eta$. We perform a similar calculation as in \cite{q6xb-8t1j} to track the steady state up to the order of $(x^2/\eta)^2$ by expanding the argument of $U_S$ to the order of $(x/\sqrt{\eta})^5$. This expansion is valid when both branches are in the and remains accurate in SRP when $\braket{x}^2g^2/\eta\ll1$. 
The resulting decoupled master equation takes the form:
\begin{eqnarray}
        \partial_t \rho' &=& \mathcal{L}'[\rho']=-i[H',\rho']+2\kappa_1\mathcal{D}_{c_1'}[\rho']+2\kappa_2\mathcal{D}_{c_2'}[\rho'],\label{eq_decoupled_6}\\
        H'&=&\frac{\Omega}{2}\sigma_z+\omega_0a^\dagger a+\frac{\omega_0}{2}\mu(a^2+{a^\dagger}^2)+\frac{\omega_0g^2}{4}\sigma_z(a+a^\dagger)^2-\frac{\omega_0 g^4}{16\eta}\sigma_z(a+a^\dagger)^4+\frac{\omega_0 g^6}{32\eta^2}\sigma_z(a+a^\dagger)^6.\nonumber           
\end{eqnarray}
The relevant terms in the transformed jump operators are obtained by expanding $\epsilon(x)$ to the leading order:
\begin{equation}
    c_1\sim a-\frac{ig}{2\sqrt\eta}\sigma_y,\,\,c_2\sim a^2-\frac{ig}{\sqrt\eta}\sigma_ya.
\end{equation}
Projecting Eq. \eqref{eq_decoupled_6} on $\ket{\pm}$ yields two decoupled master equations similar to those in \eqref*{eq_lind_eff} of the main text.  We can approximate $\rho_{ss,\pm}$ with the steady states of these equations.  
Imposing the steady-state condition $\partial_tp_+=0$ yields:
\begin{equation}
    \partial_t {\rm tr}(\ket{+}\bra{+}\rho_{ss}')={\rm tr}(\ket{+}\bra{+}\mathcal{L'}[\rho_{ss}'])=0.
    \label{eq_transformed_sstate}
\end{equation}
Explicitly evaluating Eq. \eqref{eq_transformed_sstate} gives the branch weight relation:
\begin{equation}
r_p=\frac{p_+}{p_-}=\frac{4\gamma_2n_-/\eta+\gamma_1}{4\gamma_2n_+/\eta+\gamma_1},
\label{eq_spin_pop}
\end{equation}
where $n_{\pm}={\rm tr}(a^\dagger a\rho_{ss,\pm})$. 

We verify the above formalism by numerically solving the steady state of the complete model defined in Eq. \eqref*{eq_lind} of the main text and then computing the corresponding photon number $n$ and spin imbalance $s_z$. These expectations are compared to the mixed ansatz expectations evaluated from $\rho_{ss}=U_S\rho_{ss}'U_S^\dagger$, where $\rho_{ss}'$ is constructed based on Eq.~\eqref{eq_spin_pop}. The results plotted in Fig.~\ref{fig_spin_pop} demonstrate good agreement between the two methods. 
Eq.~\eqref{eq_spin_pop} provides a straightforward explanation for the observation that the $(+)$ branch component has a much smaller weight in the composite steady state Wigner function plotted in Fig.~\ref*{fig_Wigner} of the main text, as compared to the $(-)$ branch components. $\rho_{ss,+}$ is in SRP which yields a much larger photon number compared to that of the $\rho_{ss,-}$ component near the TCP, leading to a small $p_+/p_-$ predicted by the perturbative formalism.   

The analytical formula above gives a clear and simple result, but breaks down in the regime $\braket{x}^2g^2/\eta\sim1$. A typical example would be the behavior of the QRM with only linear decay ($\gamma_2=0$) as the system is tuned away from the critical point within the SRP. In this regime $p_+/p_-$ decreases sharply as $g$ increases, resulting in a steady state $\rho_{ss}'\to\rho_{ss,-}$. This contradicts the prediction of Eq.~\eqref{eq_spin_pop} since $n_+$ remains as a small constant while $n_-$ increases dramatically. The structure of $\rho_{ss}'$ can be explained by a more precise treatment incorporating jump operators of the complete form derived in Eq. \eqref{eq_c1_comp},\eqref{eq_c2_comp}. From the MF theory, $a,a^\dagger$ can be taken as the order of $\sqrt{\eta}$, we then approximate $c_2=a^2-2i\epsilon(x)a+o(\eta^{-1})$ in the SRP because $[\epsilon(x),a]\propto\partial_x\epsilon(x)\sim o(\eta^{-1})$. Following a similar approach as before, we derive a more accurate relation for the weights of the spin branches:
\begin{equation}
    r_p=\frac{p_+}{p_-}=\frac{4\gamma_2\braket{O_2}_-/\eta+\gamma_1\braket{O_1}_-}{4\gamma_2\braket{O_2}_+/\eta+\gamma_1\braket{O_1}_+},
\label{eq_spin_pop_comp}
\end{equation}
where $O_1=\epsilon^2,O_2= a^\dagger \epsilon^2 a$ and $\braket{\dots}_{\pm}$ stand for the expectation values taken with respect to $\rho_{ss,\pm}$, the steady states of the complete decoupled Master equation  \eqref*{eq_lind_eff} of the main text. Eq.~\eqref{eq_spin_pop_comp} is a generalization of Eq.~\eqref{eq_spin_pop}. In Fig. \ref{fig_spin_pop_comp} we numerically demonstrate that it correctly captures the steady state structure when the $(-)$ branch is deep in SRP.   
\begin{figure}[t!]
    \includegraphics[width=0.75\linewidth]{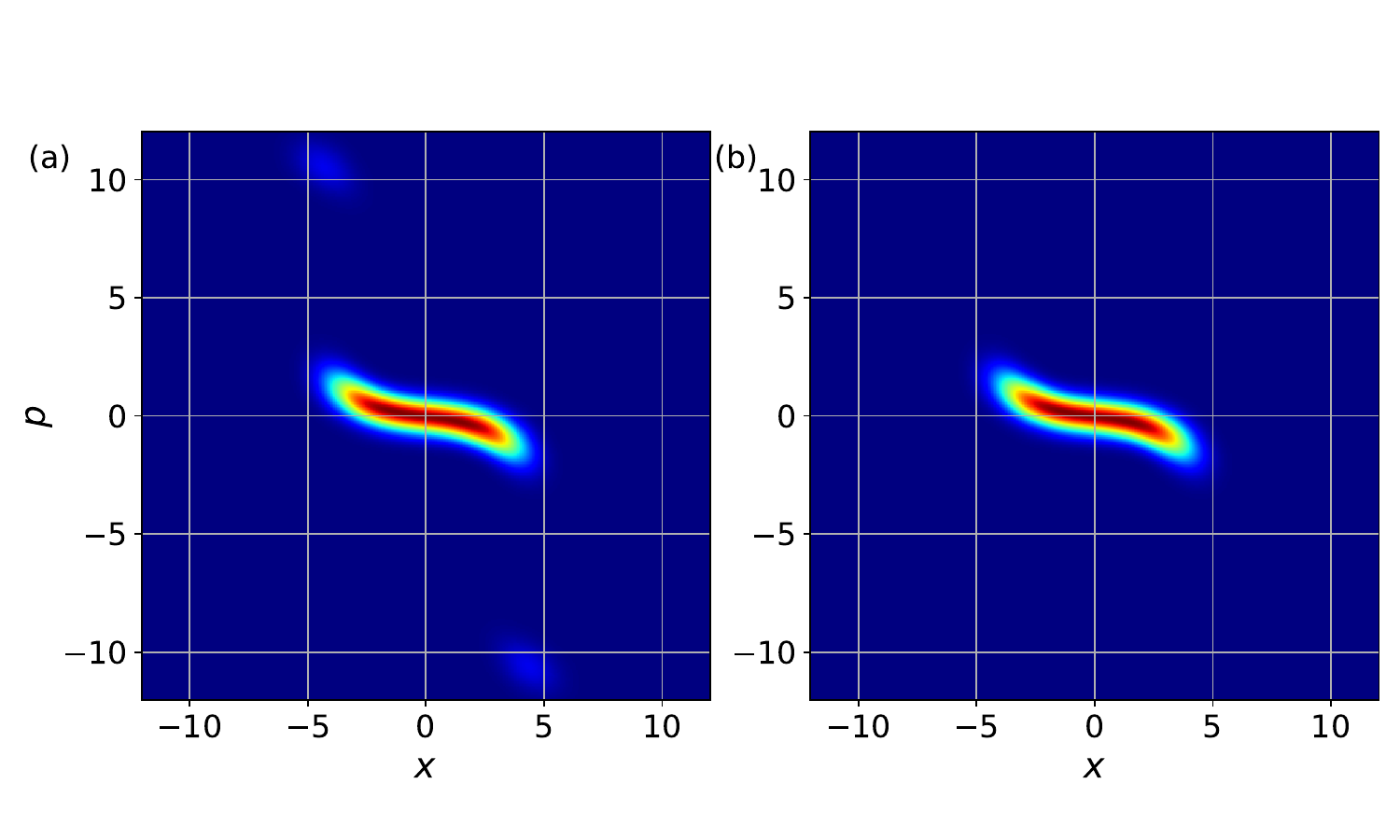}
    \caption{\small Numerical verification of the $U_{s_1}$ based rotation-measurement scheme in terms of Wigner functions at the TCP. (a). Steady-state Wigner function of the complete model (Eq. \eqref*{eq_lind}) calculated numerically. (b). Steady-state Wigner function of the $(-)$ branch, extracted using the $U_{S_1}$ based rotation-measurement scheme. Parameters: $\omega_0=1,$ $\eta=2500,g\sim1.73,\gamma_1=0.1,\gamma_2\sim44.26$. The exact values of $g,\gamma_2$ are determined by the TCP conditions in Eqs. \eqref*{eq_gc},\eqref{eq_tcp}. }
    \label{fig_Wigner_proj}
\end{figure} 
\section{Isolating the spin branches}\label{ap_iso_spin}

To isolate the spin branches, in principle, one needs to apply the complete spin rotation $U_S = \exp(-i\sigma_y/2\arctan{\left(g(a+a^\dagger\right)/\sqrt\eta)})$, as derived in the previous section. However, this transformation is difficult to realize in actual experiments. When the system is in NP or near the critical point where $gx/\sqrt{\eta}\ll1$, it is sufficient to expand the position-dependent rotation angle to the order of $1/\sqrt{\eta}$, resulting in an experimentally feasible rotation $U_{S_1}=\exp(\frac{-ig}{2\sqrt{\eta}}(a+a^\dagger)\sigma_y)$. 

We verify the effectiveness of $U_{S_1}$ numerically by applying the rotation-measurement scheme on the steady state $\rho_{ss}$ of the complete model and compare the Wigner function of the extracted state with that of $\rho_{ss}$. The results plotted in Fig.~\ref{fig_Wigner_proj} show that the $U_{S_1}$-based scheme successfully extracts the $(-)$ branch component from the steady-state mixture. 

\begin{figure}[h]
    \includegraphics[width=0.6\linewidth]{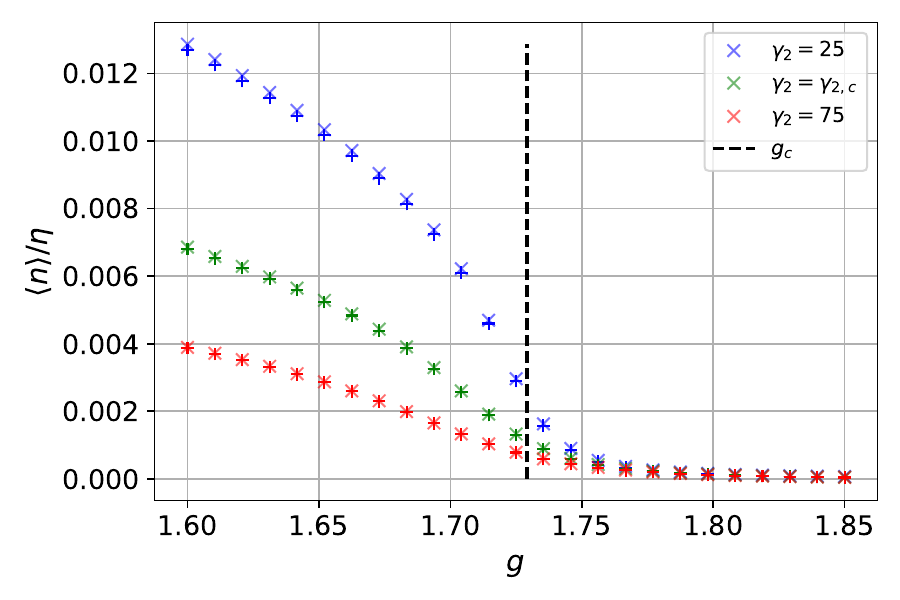}
    \caption{\small Normalized order parameter $\braket{n}/\eta$ as a function of $g$ for different $\gamma_2$ obtained with two different methods. The ``x" points of fainter colors represent the observable values extracted using the $U_{S_1}$-based scheme from the steady state of the complete model, while the ``+" points correspond to those evaluated using the decoupled master equation. Different colors denote the nature of the DPT: first-order (blue), tricritical (green), and second-order (red). The black dashed line marks the critical coupling for the second-order DPT. The parameters used for simulations are: $\omega_0=1,\eta=2500,\mu=2,$ $\gamma_1=0.1$ and the corresponding $\gamma_{2,c}\sim44.26$.
    }
    \label{fig_verify_rot}
\end{figure}

We further check if the scheme can reproduce observable values consistent with those obtained from the $(-)$ branch decoupled master equation by performing a similar set of numerical simulations as in Fig.~\ref{fig_dpt_sig}. For fixed $\omega_0,\mu,\gamma_1$ and a chosen set of $g,\gamma_2$, we first solve the complete model numerically (Eq. \eqref*{eq_lind} in the main text) and calculate $n_{ss,-}$ using $\rho_{ss,-}$ extracted via $U_{S_1}$. In Fig. \ref{fig_verify_rot} we compare the results with those obtained by directly solving the decoupled master equation (Eq. \eqref*{eq_lind_eff} of the main text). The agreement between the two sets of values validates the $U_{s_1}$-based extraction scheme. This equivalence also provides us with a resource efficient method for simulating the critical behaviors of the $(-)$ branch near the TCP.
By halving the effective Hilbert space, solving the steady state of the decoupled master equations requires only $1/4$ of the memory compared to solving the complete master equation. We leverage this computational advantage in the main text to access the large-$\eta$ regime necessary for verifying the finite-frequency scaling laws.

\section{Derivation of $W_{ss}$ using the semi-classical Langevin theory} \label{ap_langevin_derive} In this section we derive the approximated Wigner function $W_{ss}^{-}$ in Eq. \eqref*{eq_W_ss} of the main text. We focus on the $(-)$ branch while the $(+)$ branch result can be derived in a similar way. To avoid confusion, the $\pm$ labels for the branches will be dropped. We start from the decoupled master equation \eqref*{eq_lind_eff} of the $(-)$ branch. The corresponding Keldysh action is given by$S=S_F+S_D+S_I$ and
\begin{eqnarray}
S_F&=&\int{dt\alpha_+^\ast\left(i\partial_t-\omega_0\right)\alpha_+-\alpha_-^\ast\left(i\partial_t-\omega_0\right)\alpha_-},\;\;S_I=-\int d t(V_n^+-V_n^-)\nonumber\\
    S_D &=& \int dt[ - i\kappa_1( \alpha_{+} \alpha_{-}^{*} - \alpha_{+}^{*}\alpha_{+} - \alpha_{-}^{*} \alpha_{-})
- i\kappa_2  ( 
    2 \alpha_{+}^{2} \alpha_{-}^{*2} 
    - \alpha_{+}^{*2} \alpha_{+}^{2} - \alpha_{-}^{*2} \alpha_{-}^{2} )],
\end{eqnarray}
with $V_n^\pm=-\omega_0\eta/2\sqrt{g^2 (\alpha_\pm+\alpha_\pm^*)^2/\eta+1}$ and $\alpha_+,\alpha_-$ being the forward and backward fields, respectively. Performing the Keldysh rotation gives the action  
in terms of classical and quantum fields $\alpha_\pm=(\alpha_c\pm\alpha_q)/\sqrt{2}$: 
\begin{eqnarray}
    S&=&\int{dt\left[(\alpha_q^\ast\left(i\partial_t-\omega_0+i\kappa_1\right)\alpha_c+{\rm c.c.})+2i\kappa_1\alpha_q\alpha_q^\ast-\mu\left(\alpha_c\alpha_q+\alpha_c^\ast\alpha_q^\ast\right)+D_2\right]}+S_I\nonumber\\
    D_2&=&-i\kappa_2(\alpha_c\alpha_q({\alpha_q^\ast}^2+{\alpha_c^\ast}^2)-\alpha_c^\ast\alpha_q^\ast(\alpha_c^2+4\alpha_c\alpha_q+\alpha_q^2)).
    \label{eq_Keldysh_Sr}
\end{eqnarray}
We have kept the form of $S_I$ unchanged as it requires special treatment in further derivations. We study fluctuations beyond mean-field by expanding the fields around the saddle point solution $\alpha_{c,q}=\alpha_{c,q\;0}+\delta\alpha_{c,q}$. Near the TCP we have $\alpha_{c,q\;0}\rightarrow0$, the effective action $S_c$ is obtained by replacing $\alpha_{c,q}$ with $\delta \alpha_{c,q}$.
We then apply the semi-classical approximation by neglecting terms of higher than quadratic order in $\delta\alpha_q,\delta\alpha_q^*$, as they correspond to short-scale fluctuations \cite{Sieberer_2016}. For convenience, we henceforth use $\alpha_{c,q}$ to denote these fluctuations. Defining $x_{c,q}=(\alpha_{c,q}+\alpha_{c,q}^*)/\sqrt{2}$ we have 
\begin{eqnarray}
V_n^+-V_n^-\nonumber&=& - \frac{\omega_0\eta}{2}(\sqrt{g^2 (x_c+x_q)^2/\eta+1}-\sqrt{g^2 (x_c-x_q)^2/\eta+1})\nonumber\\ 
&=&- \sqrt{2}f(x_c)x_q+o(x_q^3),\;\;f(x_c)=\frac{1}{\sqrt{2}}\frac{\omega_0g^2}{\sqrt{g^2x_c^2/\eta+1}} x_c.
\end{eqnarray}

The partition function $Z=\int{\mathcal{D}[\alpha_c,\alpha_q,\alpha_c^\ast,\alpha_q^\ast]e^{iS_c}}$ is given by:
{\small
\begin{eqnarray}
     &&Z=\int\{\mathcal{D}[\alpha_c,\alpha_q,\alpha_c^\ast,\alpha_q^\ast] \nonumber\\
      &&\times \exp(\int_t{\left[\left(\partial_t-i\omega_0+\kappa_1+\kappa_2\left|\alpha_c\right|^2\right)\alpha_c^\ast-i\mu\alpha_c+if(x_c)\right]\alpha_q})\nonumber\\
      &&\times \exp(\int_t{\left[\left(-\partial_t-i\omega_0 -\kappa_1-\kappa_2\left|\alpha_c\right|^2\right)\alpha_c-i\mu\alpha_c^\ast+if(x_c)\right]\alpha_q^\ast})\nonumber\\
      &&\times \exp(\int_t{-2(\kappa_1+2\kappa_2|\alpha_c|^2) \alpha_q\alpha_q^*})\}.
\end{eqnarray}
}
The quadratic term $-2(\kappa_1+2\kappa_2|\alpha_c|^2) \alpha_q\alpha_q^*$ in the exponential can be replaced by linear terms of $\alpha_q,\alpha_q^*$ with the Hubbard-Stratonovich identity:
\begin{eqnarray}
        \exp (\int_t{-\gamma  \alpha_q\alpha_q^* })
        =\int{\mathcal{D}[\xi,\xi^\ast]\exp(\int_t{\left[-\xi^*\xi-i\sqrt{\gamma}(\alpha_q^\ast\xi-\xi^\ast\alpha_q)\right]})},
\end{eqnarray} 
where $\xi$ is a complex Gaussian white noise. Since $\kappa_1$ and $\kappa_2$ represent the rates of two independent dissipation channels, we apply the above identity to the corresponding terms separately by introducing two sets of stochastic variables $\xi_{1},\xi_{2}$. For the part with $\kappa_2$, the linear term inside the exponential is given by $\phi_2=-i2\sqrt{\kappa_2}|\alpha_c|(\alpha_q^*\xi_2-\xi_2^*\alpha_q)$. We manually add a phase factor $\theta(t)=\arg(\alpha_c)$ into the noise and define $\xi_2'=\xi_2 e^{i\theta}$ such that $\phi_2=-i2\sqrt{\kappa_2}(\alpha_c^*\alpha_q^*\xi_2'-{\xi_2'}^*\alpha_c\alpha_q)$. Here $\xi_2'$ is also a Gaussian white noise given the property of $\xi_2$. This transformation makes the resulting Langevin equation take the same form as the mean field equation when the stochastic noise terms are ignored. With the above transformations, we have:
{\small \begin{eqnarray}
        Z=&&\int  \{\mathcal{D}[\alpha_c,\alpha_c^\ast,\xi^\ast_1,\xi^\ast_1,\xi_2,\xi_2^\ast] \exp(\int_t{-(\xi_1^*\xi_1+\xi_2^*\xi_2)})\nonumber\\
    &&\times \delta[(\partial_t-i\omega_0+\kappa_1+\kappa_2\left|\alpha_c\right|^2)\alpha_c^\ast-i\mu\alpha_c+if(x_c)+i\tilde{\xi}^\ast]\ \nonumber\\
    &&\times \delta [(-\partial_t-i\omega_0 -\kappa_1-\kappa_2\left|\alpha_c\right|^2)\alpha_c-i\mu\alpha_c^\ast+if(x_c)-i\tilde{\xi}]
    \},\nonumber\\
    \tilde{\xi}=&&\sqrt{2\kappa_1}\xi_1+2\sqrt{\kappa_2}\xi_2\alpha_c^\ast,
\end{eqnarray}}
which shows that the dynamics of the system can be described by the Langevin equation:
\begin{eqnarray}
        \partial_t\alpha_c&=&\left(-i\omega_0-\kappa_1\right)\alpha_c-i\mu\alpha_c^\ast+if(x_c)-\kappa_2\alpha_c^\ast\alpha_c^2-i\sqrt{2\kappa_1}\xi_1\left(t\right)-2i\sqrt{\kappa_2\ }\alpha_c^\ast\xi_2\left(t\right).
\end{eqnarray}
With $\alpha_c=\sqrt{2}\alpha$ and $\xi_{i}=\frac{1}{\sqrt{2}}(\xi_{i,x}+i\xi_{i,p})$, we can rewrite the above Langevin equation in terms of position and momentum variables $\alpha=\frac{1}{\sqrt{2}}(x+ip)$:
\begin{eqnarray}
        \partial_tx&=&-\omega_0(\mu-1)p-\kappa_1x-\kappa_2(x^2+p^2)x+\sqrt{\kappa_1}\xi_{1,p}+\sqrt{2\kappa_2}\left(x\xi_{2,p}-p\xi_{2,x}\right),\label{eq_Langevin_xp}\\
        \partial_tp&=&-\omega_0(\mu+1)x+\frac{\omega_{0}g^{2}x}{\sqrt{2g^{2}x^{2}/\eta+1}}-\kappa_1p-\kappa_2\left(x^2+p^2\right)p\nonumber-\sqrt{\kappa_1}\xi_{1,x}-\sqrt{2\kappa_2}\left(x\xi_{2,x}+p\xi_{2,p}\right),\nonumber 
\end{eqnarray}
where $\xi_{i,x},\xi_{i,p}$ are independently distributed Gaussian white noises which satisfy $\braket{\xi_{i,x}(t)\xi_{j,p}(t')}=0,\; \braket{\xi_{i,x}(t)\xi_{j,x}(t')}=\braket{\xi_{i,p}(t)\xi_{j,p}(t')}=\delta_{ij}\delta(t-t')$. We note that \cite{PhysRevResearch.7.013061} studies a bosonic Hamiltonian that is equivalent to our $H_-$ with $V_{\rm nl}$ expanding to the order $(a+a^\dagger)^2/\eta$. A similar set of Langevin equations is derived without the one-photon decay using the Fokker-Planck equation of the
truncated Wigner function. 

To obtain the analytical steady state solution for  \eqref{eq_Langevin_xp}, we follow a similar approach as in \cite{PhysRevResearch.7.013061} by considering the limit where $\Delta x^2\gg$  $\Delta p^2$. The relative magnitude of these fluctuations can be estimated using the mean-field equation \eqref{eq_fixed_point_u}. Near the critical point we assume $u_p,u_x$ are small such that the term with coefficient $\gamma_2$ can be neglected, resulting in 
\begin{equation}
    r_{f}=u_x/u_p\sim(1-\mu)^2/\gamma_1^2.
    \label{eq_x_p_ratio}
\end{equation}
With $r_f\gg1$ one can follow the same scaling analysis presented in the appendix of Ref.~\cite{PhysRevResearch.7.013061} to show that, the stochastic terms $x\xi_{2,p},p\xi_{2,x}$ in the $x$-equation, $p\xi_{2,p}$ in the $p$-equation, and all terms containing $p^2$ in Eq.~\eqref{eq_Langevin_xp} can be neglected when $\eta$ becomes sufficiently large. Under these approximations, Eq.~\eqref{eq_Langevin_xp} reduces to:
\begin{eqnarray}
    \partial_tx&=&-\omega_0(\mu-1)p-\kappa_1x-\kappa_2x^3+\sqrt{\kappa_1}\xi_{1,p},\nonumber\\
    \partial_tp&=&-\omega_0(\mu+1)x+\frac{\omega_{0}g^{2}x}{\sqrt{2g^{2}x^{2}/\eta+1}}-\kappa_1p-\kappa_2x^2p-\sqrt{\kappa_1}\xi_{1,x}-\sqrt{2\kappa_2}x\xi_{2,x}.
\end{eqnarray}
Since we have assumed $\omega_0^2(\mu-1)^2\gg\kappa_1^2$, the fluctuations in $x$ are mainly induced by the momentum through the dominant driving term $-\omega_0(\mu-1)p$, such that the stochastic noise $\sqrt{\kappa_1}\xi_{1,p}$ in the $x$-equation can be neglected. Defining $v=\partial_tx$ and substituting the $p$-equation into $\partial_tv$ gives:
\begin{eqnarray}
    \partial_t v+(2\kappa_1+4\kappa_2x^2)v=-\frac{1}{m}\partial_xU\left(x\right)
    +{\rm sgn}(T_{\rm eff})2\sqrt{T_{\rm eff}/m}(\sqrt{\kappa_1}\xi_{1,x} + \sqrt{2\kappa_2}x\xi_{2,x})  \label{eq_langevin_xv}
\end{eqnarray}
where $U,T_{\rm eff},v$ are defined as in Eq. \eqref*{eq_eff_variables} of the main text. The above stochastic equation is then equivalent to the Fokker-Planck equation of the Wigner distribution \cite{carmichael2013statistical,gardiner2004handbook}: 
\begin{eqnarray}
    \partial_tW=\left[-\partial_xv+\partial_v(\frac{1}{m}\partial_xU+2\left(\kappa_1+2\kappa_2x^2\right)v)\right]W+\frac{T_{\rm eff}}{m}2\left(\kappa_1+2\kappa_2x^2\right)\partial_v^2W,
    \label{eq_FP}
\end{eqnarray}
which yields a steady state solution in the Boltzmann form given by Eq. \eqref*{eq_W_ss} in the main text.

\begin{figure*}[t!]
    \centering
    \includegraphics[width=0.75\textwidth]{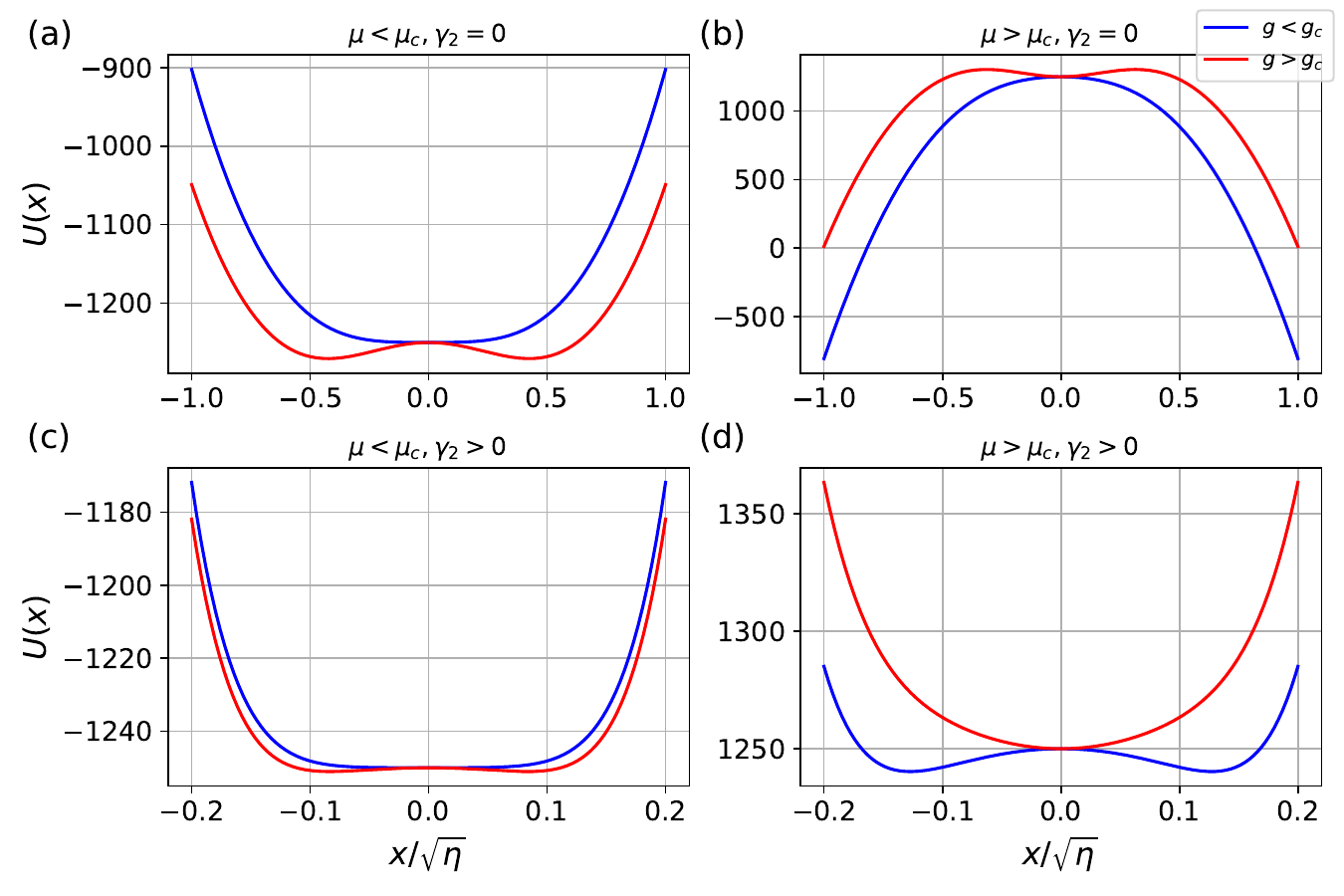}
    \caption{\small Effective potential $U(x)$ of the $(-)$ branch as a function of normalized coordinate $x/\sqrt{\eta}$ for different regimes. \textbf{(a,b)} are plotted without nonlinear dissipation ($\gamma_2=0$) while \textbf{(c,d)}  assumes $\gamma_2=50$. \textbf{(a,c)} represent the ordinary regime $\mu=0<\mu_c$ while\textbf{(b,d)} represent the inverted regime $\mu=2>\mu_c$. The blue and red solid curves stand for the potential at $g<g_c$ and $g>g_c$. For the other parameters in the model we have used $\omega_0=1,\eta=2500,\gamma_1=0.1$. In the ordinary regime we take $g=1,1.1$ while in the inverted regime we choose $g=1.5,2$. }
    \label{fig_U_x}
\end{figure*}

\section{DPTs in the inverted regime}\label{ap_invert}
To understand physics in the inverted regime ($\mu>\mu_c=\sqrt{1+\gamma_1^2}$), we adopt the decoupled Master equations \eqref*{eq_lind_eff} of the main text. The corresponding MF Heisenberg equations are given by:
\begin{eqnarray}
        \partial_tx&=&-\omega_0(\mu-1)p-\kappa_1x-\kappa_2(x^2+p^2)x,\nonumber\\
        \partial_tp&=&-\omega_0(\mu+1)x\mp f_{\rm nl}(x/\sqrt{\eta})-\kappa_1p-\kappa_2\left(x^2+p^2\right)p,
    \label{eq_mf_decouple}
\end{eqnarray}
where $f_{\rm nl}$ is the nonlinear force defined in the normalized MF equation for the complete model in the main text. To gain some intuitions, we first disregard the dissipation by setting $\kappa_1=\kappa_2=0,(\mu_c=1)$ such that: 
\begin{eqnarray}
        \partial_tx=-\omega_0(\mu-1)p, \;\;\nonumber\partial_tp=-\omega_0(\mu+1)x\mp f_{\rm nl}(x/\sqrt{\eta})
    \label{eq_mf_coh}
\end{eqnarray}
This equation can be interpreted as the equation of motion for a particle of effective mass $m_0=-[\omega_0(\mu-1)]^{-1}$ moving in the potential $U_0(x)=\frac{1}{2}\omega_0^2(\mu+1)x^2\pm V_{\rm nl}(x)$ with $V_{\rm nl}$ defined in Eq.~\eqref*{eq_V_nl} of the main text. When $\mu<\mu_c$, the effective mass is positive ($m_0>0$), the harmonic part of $U_0(x)$ establishes a stable minimum at $x=0$. For the $(-)$ branch, the nonlinear potential $-V_{\rm nl}$ counteracts against the harmonic confinement and eventually removes the minimum at the origin when $g$ becomes sufficiently large, leading to superradiance. In contrast, when $\mu>\mu_c$, the particle exhibits a negative effective mass. The induced dynamics is physically equivalent to a particle with mass $|m_0|$ in an inverted potential $U'_0=-U_0$, effectively reversing the roles of the harmonic and nonlinear components of the potential. In this case, $V_{\rm nl}$ generates the confining force and restores the minimum at $x=0$ when $g$ becomes sufficiently large, resulting in a transition from the SRP back to the NP. The point $\mu=\mu_c=1$ is identified as the critical point for spectral collapse \cite{Downing2023}, beyond which the Hamiltonian no longer supports bound states. We note that due to the stabilization effect of the two-photon dissipation, the model we investigate here always yields a stable steady state even for the case $g=0, \mu>\mu_c$, where the Hamiltonian reduces to two uncoupled oscillators with PA. A second-order superradiant DPT has been identified in this decoupled model in Refs.~\cite{PhysRevA.94.033841,PhysRevResearch.7.013061}. The above arguments also explain the behavior of $F_-(\bar{x})$ at different values of $g$ in Fig. \ref*{fig_phase_diag}(b) of the main text.

Now let us consider the presence of one- and two-photon dissipation channels. The effective potential $U(x)$ we derived using the Langevin formalism allows us to gain a deeper understanding of the inverted regime while incorporating the effects of dissipation. With the choice of $m=1$, we have already encoded the impact of the effective mass in the form of the potential. The critical PA strength shifts to $\mu_c=\sqrt{1+\gamma_1^2}$ as a result of the linear dissipation, which can be obtained easily by setting the coefficient of $x^2$ to $0$. In Fig. \ref{fig_U_x} we provide four example plots of $U(x)$ that cover the ordinary ($\mu<\mu_c$) and the inverted ($\mu>\mu_c$) regime, with and without two-photon decay. When $\mu<\mu_c$, a large $g$ turns the global minimum at $x=0$ into a local maximum, indicating a transition from NP to SRP. For $\mu>\mu_c, \gamma_2>0$, it does the opposite by turning the local maximum at $x=0$ into a global minimum,  indicating a transition from SRP to  NP. For $\mu>\mu_c, \gamma_2=0$, a large $g$ turns the global maximum at $x=0$ into a local minimum while the system stabilizes in NP. The comparison between Fig. \ref{fig_U_x} (b) and (d) also visualizes how the two-photon decay stabilizes the SRP in the inverted regime by introducing a high order confining potential. We note that Fig.~\ref{fig_U_x} (b),(d) are consistent with Fig.~\ref*{fig_phase_diag}(b), (c) in the main text, demonstrating the equivalence between $U(x)$ and $F(\bar{x})$ in the limit $\mu-1\gg\gamma_1$. 

\begin{figure*}[t!]
    \centering
    \includegraphics[width=0.9\textwidth]{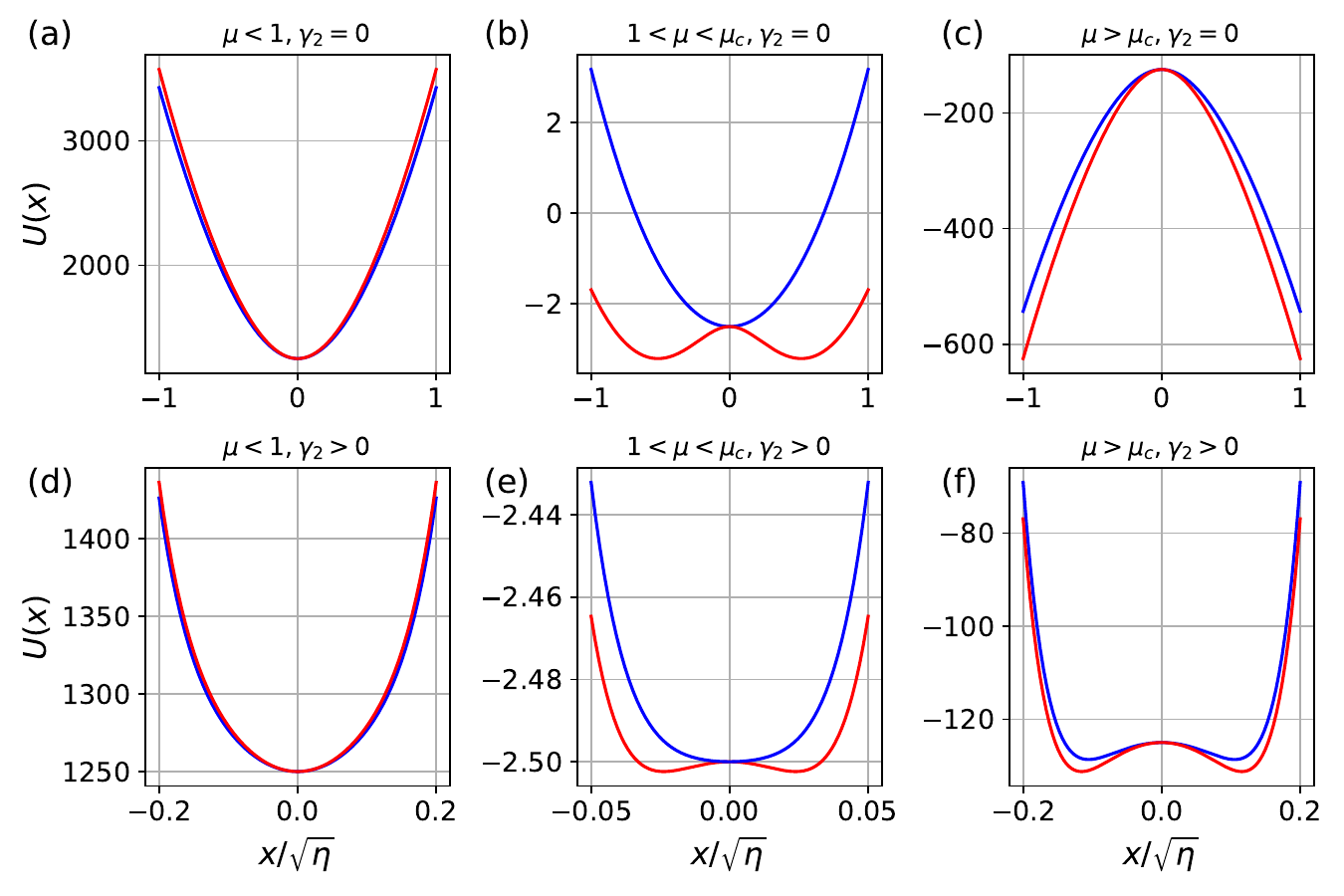}
    \caption{\small Effective potential $U(x)$ for the $(+)$ branch as a function of normalized coordinate $x/\sqrt{\eta}$ for different regimes. \textbf{(a,b,c)} are plotted without nonlinear dissipation ($\gamma_2=0$) while \textbf{(d,e,f)}  assumes $\gamma_2=50$. \textbf{(a,c),(b,d),(c,f)} represent the regimes A: $\mu=0<1$, B:$1<\mu=1.002<\mu_c$, C:$\mu=2>\mu_c$. The blue and red solid curves stand for the potential at $g<g_c$ and $g>g_c$. Here $g_c$ is the critical coupling for the $(+)$ branch in B and that for $(-)$ branch in A,C. For the other parameters in the model we have used $\omega_0=1,\eta=2500,\gamma_1=0.1$. We take $g=1,1.1$ in A, $g=1,2.5$ in B and $g=1.5,2$ in C. }
    \label{fig_U_x_up}
\end{figure*}

For reference, we have also attached example plots of $U(x)$ for the $(+)$ branch\footnote{The effective potential of the $(+)$ branch can be obtained simply by reversing the sign of $V_{\rm nl}$ in Eq. \eqref*{eq_eff_variables} of the main text} in Fig. \ref{fig_U_x_up}. In the $\mu<1$ regime, no non-trivial local minimum exists, such that the $(+)$ branch stays in the NP. When $1<\mu<\mu_c$, a pair of symmetric local minimums emerges when the spin-boson coupling is sufficiently large, corresponding to a superradiant DPT at large $g$. Finally in the inverted regime, the potential supports no minimum with $\gamma_2=0$ and always supports two stable non-trivial local minimums when $\gamma_2>0$. This is consistent with the previous calculation that the $(+)$ branch supports no stable NP or SRP with only linear decay and stays in the SRP when the non-linear decay stabilizes the system.  

As shown in Fig.~\ref*{fig_mf_n_sz} of the main text, the fundamentally different behaviors in the ordinary ($\mu<\mu_c$) and the inverted regimes are clearly reflected by the MF order parameters $\bar{n}_{ss},s_z$. The plot for $\bar{n}_{ss}$ can already be well explained by our analysis above. For the $s_z$ plot, we shall notice that the spin polarization is governed by the relation $s_z=\pm\frac{1}{\sqrt{1+2g^2\bar{x}^2}}$ (from Eq. \eqref{eq_fixed_point_complete}). In the ordinary regime , $\bar{x}$ gains a non-zero value when the spin-boson coupling is strong enough to overcome the confinement generated by the amplified harmonic potential. For $g>g_c$, $x$ grows continuously with $g$, resulting in a strictly monotonic increase (decrease) in $s_z$ for the $(-)$ ($(+)$) branch. Conversely, in the inverted regime, the spin-boson coupling acts as a confining potential for the $(-)$ branch, restricting non-zero $\bar{x}$  solutions to the weak-coupling regime ($g<g_c$). When the transition occurs, $g\bar{x}$ jumps abruptly, pushing $s_z$ sharply above $-1$. As $g$ decreases further, $\bar{x}$ varies only slightly, causing $g\bar{x}$ to shrink back towards 0. This mechanism dictates the complex, non-monotonic behavior of $s_z$ in the inverted regime, where the system ultimately mimics a return to the normal phase $(s_z\to -1)$ when $g$ vanishes.   
\section{Evaluation of steady-state observables}
\label{ap_obs_ss}
 
We can evaluate the steady-state expectation value of the symmetric-ordered bosonic operator $(p^kx^m)_S$ by integrating $p^kx^mW_{ss}$ over $p$ with $W_{ss}$ given in Eq. \eqref*{eq_W_ss} of the main text \cite{PhysRevResearch.7.013061}:
{\small\begin{equation}
    \langle p^kx^m\rangle=\left(\frac{i}{2\sqrt2}\right)^k\int_{-\infty}^{\infty}{dx\ x^mH_k[\frac{i\sqrt2}{\mu-1}(\gamma_1x+\gamma_2x^3/\eta)]W_R\left(x\right)},
    \label{eq_xp_exp}
\end{equation}}
where $H_k$ is the Hermite polynomial and $W_{R}$ is the reduced Wigner distribution:
\begin{equation}
    W_R(x)=\int_{-\infty}^{\infty} \frac{dp}{2}W_{ss}(x,p)=\frac{1}{Z_0'}\exp{\left(-U\left(x\right)/T_{\rm eff}\right)},
\end{equation}
with $Z_0'$ being a normalization constant. 
We first evaluate the critical exponent assuming $g_c\lesssim g$ but the difference $g^2-g^2_c$ is sufficiently large compared to $1/\eta$ such that $C_2\gg C_4$. We note that $g$ cannot be far away from $g_c$, otherwise the semi-classical approximation breaks down. Evaluating the integral gives:
\begin{equation}
    \braket{x^2}=\frac{\mu-1}{4(g^2-g^2_c)},\;\;\braket{p^2}\sim \frac{1}{4},\;\;\braket{n}=\frac{\mu-1}{8(g^2-g^2_c)}-\frac{3}{8}.
\end{equation}
Here $\Delta x^2=\braket{x}^2,\Delta p^2=\braket{p}^2$ since $\braket{x},\braket{p}=0$ are enforced by the weak $Z_2$ symmetry.

We then evaluate the $x,p$ quadratures using Eq. \eqref{eq_xp_exp} at the critical point $g=g_c$ such that the $x^2$ term in $U(x)$ vanishes. When $C_4>0$, the $x^4$ term dominates at large $\eta$ such that:
\begin{eqnarray}
\braket{x^2}&=&\frac{\Gamma(1/4)}{\Gamma(3/4)}\beta_1^{-1/2},\;\;
\braket{p^2}=\frac{\gamma_1^2}{(\mu-1)^2}\frac{\Gamma(1/4)}{\Gamma(3/4)}\beta_1^{-1/2}+\frac{1}{2}\nonumber\\
\braket{n}&=&\frac{1}{2}\left(1+\frac{\gamma_1^2}{(\mu-1)^2}\right)\frac{\Gamma(1/4)}{\Gamma(3/4)}\beta_1^{-1/2}-\frac{1}{4}.
\end{eqnarray}
In the  $C_4=0$ case (TCP), we have
\begin{eqnarray}
\braket{x^2}=\frac{\sqrt{\pi}}{\Gamma(1/6)}\beta_2^{-1/3}, \;\;
\braket{p^2}=\frac{1}{2},\;\;
\braket{n}=\frac{1}{2}\frac{\sqrt{\pi}}{\Gamma(1/6)}\beta_2^{-1/3}-\frac{1}{4}.
\end{eqnarray}
Here we have defined $\beta_1= C_4/(4T_{\rm eff}),\beta_2=C_6/(6T_{\rm eff})$.

\section{Higher-order multicriticality}
\label{multi_c}
Since $F(x)$ is in principle of infinite order, it is worth searching for multicritical points. We briefly investigate the existence of a tetracritical point in the $(-)$ branch. To see this, we expand $G(u_x)$ to the order of $N=2$ to obtain a complete expression of $c_6$:
{\small 
\begin{equation}
    c_6^-=\frac{1}{6(\mu-1)^2}[2(g^2-2\mu)^2\gamma_2^2+4g^4\gamma_2 \gamma_1\left(-1+\mu\right)+3g^6\left(-1+\mu\right)^3].
\end{equation}
}
The tetracritical point yields $c_2^-=c_4^-=c_6^-=0$. Using $c_2^-=c_4^-=0$ we can solve for $\gamma_1,\gamma_2$ in terms of $\mu,g$. Substituting into to the equation $c_6^-=0$ gives:
\begin{equation}
g^2=6\mu+1\pm\sqrt{24\mu^2+1}.
\end{equation}
For $\mu>1$, $g^2>2$, the critical condition requires $\gamma_1^2=(\mu-1)(-5\mu+\sqrt{24\mu+1})<0$ which is impossible. 

A similar conclusion can be reached by examining the effective potential $U(x)$, as the sixth order coefficient $C_6$ is strictly positive in the inverted regime. It can also be observed that the two-photon decay will not further affect higher order terms in the potential, as the highest order term that contains $\gamma_2$ is $x^6$. To manipulate multicriticalities beyond tricriticallity, coherent and dissipative processes of higher-order need to be introduced, such as three-photon interaction/decay.
\bibliography{refs}{}